\documentclass[reprint, superscriptaddress, amsmath,amssymb, aps,
prb]{revtex4-1}
\pdfoutput=1
\usepackage{graphicx}
\usepackage{dcolumn}
\usepackage{bm}
\usepackage{xcolor}
\usepackage[normalem]{ulem}
\usepackage{hyperref}

\newcommand{\ii} {\mathrm{i}}
\newcommand{\ee} {\mathrm{e}}
\newcommand{\vep} {\varepsilon}
\newcommand{\dgr} {^{\circ}}

\newcommand{\ind}[1] {{\mathrm{#1}}}
\newcommand{\vect}[1] {\boldsymbol{\mathbf{#1}}}
\newcommand{\vers}[1] {\boldsymbol{\mathbf{\hat{#1}}}}
\newcommand{\unit}[1] {\mbox{ }\mathrm{#1}}

\DeclareMathOperator{\arccosh}{arccosh}

\begin{document}


\title{Probing Nonlocal Effects in Metals with Graphene Plasmons}

\author{Eduardo J. C. Dias}
 \email{eduardo.dias@fisica.uminho.pt}
\affiliation{%
 Department of Physics and Center of Physics, and QuantaLab, University of Minho, PT--4710--057, Braga, Portugal
}%
\author{David Alcaraz Iranzo}
\affiliation{ICFO - The Institute of Photonic Sciences, The Barcelona Institute of Science and Technology, 08860 Castelldefels (Barcelona), Spain}
\author{P.~A.~D.~Gon\c{c}alves}
\affiliation{Center for Nano Optics, University of Southern Denmark, Campusvej 55, DK-5230~Odense~M, Denmark}
\affiliation{Department for Photonics Engineering, Technical University of Denmark, DK-2800 Kongens Lyngby, Denmark}
\affiliation{Center for Nanostructured Graphene, Technical University of Denmark, DK-2800 Kongens Lyngby, Denmark}
\author{Yaser Hajati}
\affiliation{%
  Department of Physics, Faculty of Science, Shahid Chamran University of Ahvaz, Ahvaz, Iran
}%
\author{Yuliy V. Bludov}
\affiliation{%
	Department of Physics and Center of Physics, and QuantaLab, University of Minho, PT--4710--057, Braga, Portugal
}%
\author{Antti-Pekka Jauho}
\affiliation{Department of Micro- and Nanotechnology, Technical University of Denmark, DK-2800 Kongens Lyngby, Denmark}
\affiliation{Center for Nanostructured Graphene, Technical University of Denmark, DK-2800 Kongens Lyngby, Denmark}
\author{N. Asger Mortensen}
\affiliation{Center for Nano Optics, University of Southern Denmark, Campusvej 55, DK-5230~Odense~M, Denmark}
\affiliation{Danish Institute for Advanced Study, University of Southern Denmark, Campusvej 55, DK-5230~Odense~M, Denmark}
\affiliation{Center for Nanostructured Graphene, Technical University of Denmark, DK-2800 Kongens Lyngby, Denmark}

\author{Frank H. L. Koppens}
\affiliation{ICFO - The Institute of Photonic Sciences, The Barcelona Institute of Science and Technology, 08860 Castelldefels (Barcelona), Spain}
\affiliation{ICREA - Instituci\'{o} Catalana de Recerça i Estudis Avan\c{c}ats, Barcelona, Spain
}%
\author{N. M. R. Peres}%
\email{peres@fisica.uminho.pt}
\affiliation{%
	Department of Physics and Center of Physics, and QuantaLab, University of Minho, PT--4710--057, Braga, Portugal
}%
 

\date{\today}

\begin{abstract}
In this paper we analyze the effects of nonlocality on the optical properties of a system consisting of a thin metallic film separated from a graphene sheet by a hexagonal boron nitride (hBN) layer. We show that nonlocal effects in the metal have a strong impact on the spectrum of the surface plasmon-polaritons on graphene. If the graphene sheet is shaped into a grating, we show that the extinction curves can be used to shed light on the importance of nonlocal effects in metals. Therefore, graphene surface plasmons emerge as a tool for probing nonlocal effects in metallic nanostructures, including thin metallic films. As a byproduct of our study, we show that nonlocal effects lead to smaller losses for the graphene plasmons than what is predicted by a local calculation. We show that these effects can be very well mimicked using a local theory with an effective spacer thickness larger than its actual value.
\end{abstract}

\maketitle


\section{\label{sec:introduction}Introduction}

Nanoplasmonics is a field of optics dealing with the interaction of electromagnetic radiation with metallic nanostructures and nanoparticles\cite{Maier-book,book-nanoparticles,goncalves2016introduction,nanoplasmonics-book}. Over the last decade, the characteristic size of plasmonic structures has been steadily  approaching the few-nanometer scale~\citep{fernandez2017unrelenting}, with concomitantly ultra-confined plasmonic modes~\citep{benz2016single,ciraci2012probing,Raza:2015b,lundeberg2017tuning}. 
The most common description of the electrodynamics governing plasmonic systems typically ignores that the response of a metallic nanostructure is controlled by a nonlocal dielectric tensor \cite{Mortensen-2011,Mortensen-2014}. Indeed, the general linear-response expression for the electric displacement vector reads~\citep{Jackson}
\begin{equation}
\mathbf{D}(\mathbf{r},\omega)=\vep_0\int d\mathbf{r}^\prime\bar{\bm \vep}(\mathbf{r},\mathbf{r}^\prime,\omega) \mathbf{E}(\mathbf{r}^\prime,\omega),
\label{eq_nonlocal_real_space}
\end{equation}
where $\mathbf{D}(\mathbf{r},\omega)$ is the electric displacement vector, $\mathbf{E}(\mathbf{r},\omega)$ is the electric field, and $\bar{\bm \vep}(\mathbf{r},\mathbf{r}^\prime,\omega)$ is the nonlocal dielectric tensor. For a translational invariant system, we have $\bar{\bm \vep}(\mathbf{r},\mathbf{r}^\prime,\omega)=\bar{\bm \vep}(\mathbf{r}-\mathbf{r}^\prime,\omega)$. Equation~(\ref{eq_nonlocal_real_space}) embodies the statement that the electric displacement field at point $\mathbf{r}$ depends on the electric field at all points $\mathbf{r}^\prime$. 

In general, and in particular for systems with translational invariance, it is convenient to transform Eq.~\eqref{eq_nonlocal_real_space} to momentum space, obtaining
$\mathbf{D}(\mathbf{k},\omega)=\vep_0\bar{\bm \vep}(\mathbf{k},\omega)\mathbf{E}(\mathbf{k},\omega)$,
where translational invariance has been assumed and $\mathbf{k}$ denotes the wave vector. The local response approximation (LRA) is equivalent to neglecting the wave vector dependence of the dielectric tensor by taking the long-wavelength limit ($\mathbf{k} \rightarrow 0$). However, 
when the system's characteristic length scales approach the nanometer range, the wave vector dependence of the dielectric tensor has profound consequences on the spectrum of plasmonic resonances. It follows from Maxwell's equations that the wave equation in momentum space reads~\cite{goncalves2016introduction}
%
\begin{align}
-\vect{k} \times \left[ \vect{k} \times \vect{E}(\vect{k},\omega) \right] &= -\vect{k} \left[ \vect{k} \cdot \vect{E}(\mathbf{k},\omega) \right] +k^2 \mathbf{E}(\mathbf{k},\omega) 
\nonumber
\\ &=\frac{\omega^2}{c^2}\bar{\bm \vep}(\mathbf{k},\omega)\mathbf{E}(\mathbf{k},\omega).
\label{eq:2}
\end{align}
Equation (\ref{eq:2}) has two types of solutions. The divergence-free solution 
($\vect{k} \cdot \vect{E} = 0$)
corresponds to the usual wave equation $k^2 \vect{E} = \bar{\bm \vep} (\omega/c)^2 \vect{E}$; in this context, this solution is usually dubbed as the transverse mode. However, within the nonlocal framework, there is an additional curl-free solution ($\vect{k} \times \vect{E} = \vect{0}$) which, for a given frequency $\omega$, is obtained when the condition $\vep(\vect{k},\omega) = 0$ is satisfied. This solution is traditionally referred to as the longitudinal mode, although one should bear in mind that this mode is also perpendicular to the direction of propagation of the field.

\begin{figure}[htbp]
 \centering
 \includegraphics[width=\linewidth]{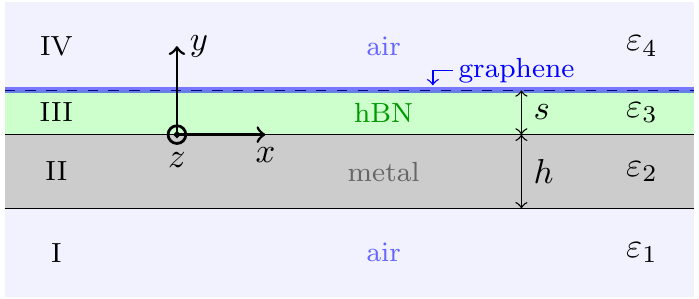}
 \caption{Schematic representation of one type of layered heterostructure investigated in this work. We assume the system to be invariant in the $x$-- and $z$--directions.}
 \label{fig:scheme:4layer}
\end{figure}

The longitudinal solution is generally overlooked within the LRA, which may reveal itself as
an inaccurate
approximation when either the size of the metallic nanostructures or the separation between 
two metallic surfaces fall below a couple of tens of nanometers.~\cite{Raza:2013,david2011spatial,ciraci2012probing}. Here, we consider a geometry 
similar to the latter, namely a configuration in which a graphene sheet is placed parallel to and 
extremely close to a metal surface---see Fig.~\ref{fig:scheme:4layer}.
The spectrum of the aforementioned structures can be computed either analytically or numerically---depending of the geometry---and such calculations show that the resonances associated with the excitation of surface plasmon-polaritons (either localized or propagating) exhibit strong deviations from the ones obtained within the LRA~\cite{fernandez2012transformation,scholl2012quantum}. Additionally, the local-response calculation predicts field-enhancements which are larger than in the nonlocal case~\cite{Khurgin:2017,Garcia-de-Abajo:2008aa,David:2010aa}. This is particularity true when we have two nanoparticles in close proximity~\cite{Khurgin:2017,Wiener:2013aa,book-nanoparticles,Schnitzer:2016aa,toscano2012modified}. 
Experimentally, one can investigate nonlocal effects using diverse techniques, such as electron energy-loss spectroscopy~\cite{Wiener:2013aa,Raza:2015b} and far-field spectroscopy \cite{Shen:2017aa}. 
Therefore, a proper theoretical description of nonlocal plasmonics requires the calculation 
of a suitable nonlocal dielectric function (or, equivalently, the nonlocal conductivity). 
A possible approach is using {\it ab-initio} methods~\cite{Vidal:2013,Alejandro:2016aa}. However, typical (albeit small) plasmonic structures involve a large number of atoms which render those methods difficult to use in a routinely fashion, since they quickly become time-consuming and computationally demanding.
Hence, it is natural to seek for an alternative approach to the calculation of the nonlocal dielectric function. As early as in the 1970s, it was observed that the plasmonic properties of thin metallic films did not follow the prediction of the local-response theory~\cite{Boardman:1975aa,boardman1982electromagnetic}. The need to reconcile theory and experiment required a description of an electron gas subjected to external fields introduced by Bloch\cite{bloch1933bremsung,bloch1933bremsvermogen} and Jensen\cite{jensen1937eigenschwingungen}. This model of an electronic fluid came to be known as the hydrodynamic model\cite{boardman1982electromagnetic,moreau2013impact,ciraci2013hydrodynamic,Raza:2013,Raza:2015}. This model combines Maxwell's equations with Newton's second law of motion for a charged particle supplemented with a term taking into account the statistical pressure in the electron gas due to charge inhomogeneity~\cite{Boardman:1975aa,boardman1982electromagnetic,Raza:2013}. 
The revival of plasmonics unburied the hydrodynamic model and applied it to the optical response of metallic nanoparticles with great success~\cite{ciraci2012probing,ciraci2013hydrodynamic,toscano2012modified}. As already mentioned, both the position and width of the plasmon resonance, and the field enhancement measured experimentally do not agree with the approach using the LRA, whereas the hydrodynamic model is able to explain the experimental results~\cite{ciraci2012probing}. For this reason, the hydrodynamic model has now become a popular approach for nonlocal optics~\cite{Mortensen-2014,Toscano:2015aa,Raza:2013}.

More recently, the birth of 2D crystals\cite{Novoselov:2005aa} introduced another possibility for studying nonlocal effects in the optical response of materials. In particular, the 
emergence of graphene plasmonics~\cite{goncalves2016introduction} constitutes a new playground for studying these effects down to the one atom thick limit~\cite{woessner2015highly,lundeberg2017tuning}. The combination of graphene with two-dimensional (2D) insulators---such as hexagonal Boron Nitride (hBN)---has allowed an unprecedented control of the distance that a 2D material can be placed in the vicinity of metallic films or nanostructures. In fact, the placement of a graphene sheet at a distance of a few nanometers away from a metal surface has recently been experimentally demonstrated~\cite{lundeberg2017tuning}, in a similar setting as depicted in Fig.~\ref{fig:scheme:4layer}. This geometry provides a suitable way to obtain deep 
subwavelength confinements down to the atomic limit. This is possible due to the interplay of graphene plasmons and the screening exerted by the nearby metal film. This approach has revealed that the nonlocal properties of graphene's conductivity have to be included in a proper account of the experimental data~\cite{lundeberg2017tuning,principi2014plasmon}. 
In graphene, it is more useful to describe the nonlocal response via the material's nonlocal dynamical conductivity, which within linear-response theory relates the surface current, $\mathbf{K}_\ind{s}(\mathbf{r},\omega)$, and the in-plane electric field, $\mathbf{E}_\parallel(\mathbf{r},\omega)$, via~\cite{goncalves2016introduction} 
\begin{equation}
	\mathbf{K}_\ind{s}(\mathbf{r},\omega)=\int d\mathbf{r}^\prime\bar{\bm \sigma}(\mathbf{r}-\mathbf{r}^\prime,\omega) \mathbf{E}_\parallel(\mathbf{r}^\prime,\omega).
	\label{eq_nonlocal_real_space_sigma}
\end{equation}
Notice that this statement is merely a reformulation of Eq.~(\ref{eq_nonlocal_real_space}). Here, $\bar{\bm \sigma}(\mathbf{r},\omega)$ refers to the nonlocal (surface) conductivity tensor of graphene. This quantity can be probed experimentally using graphene plasmons when the material is in the proximity of a metallic surface, and varying the graphene-metal distance, thereby retrieving the Fourier transform of $\bar{\bm \sigma}(\mathbf{r},\omega)$, 
$\bar{\bm \sigma}(\mathbf{q},\omega)$~\cite{lundeberg2017tuning}. In the absence of strain, graphene is isotropic and thus $\bar{\bm \sigma}(\mathbf{q},\omega)=\bar{\bm \sigma}(q,\omega)$, where $\mathbf{q}$ labels the in-plane wave vector.

If the graphene sheet is transformed into a grating (or an external grating is deposited on the graphene sheet), the properties of the plasmons depend sensitively on the lenght of the imposed period~\cite{goncalves2016introduction,gonccalves2016modeling,dias2017controlling}. 
Therefore, adjusting the grating period can be used to tune the graphene plasmons' frequency to the desired spectral range. Either graphene itself~\cite{gonccalves2016modeling} or a grating may be patterned~\cite{dias2017controlling}. 
When graphene is placed in the vicinity of a metallic surface, e.g., as illustrated in Fig.~\ref{fig:scheme:4layer}, nonlocal effects arising from the metal's response may influence indirectly the behavior of the fields in the region between graphene and the metal, namely through a larger penetration of the field inside the metal due to a less effective screening induced by nonlocality. The quantification of the importance of such nonlocal effects constitutes the goal of this work.

Here, we investigate the influence of the nonlocal response of a metal on the optical response of systems based on graphene lying in close proximity to a metallic film or surface. In what follows, we consider both the case in which the metal is constituted of Gold and of Titanium. 
We shall focus on the latter, as this metal exhibits a very strong nonlocal behavior (compared, for example, with Gold). The nonlocal effects arising from the metal's response are evaluated both on the plasmonic (near-field physics) and on the optical properties (far-field spectroscopy) of these structures. Throughout the manuscript, graphene's conductivity is taken as being nonlocal~\cite{goncalves2016introduction}. The aim of this work is therefore to use graphene plasmonics to probe nonlocal effects within the metal, in opposition to the study of nonlocality in graphene performed in a recent publication~\cite{lundeberg2017tuning}.

Specifically, we study the impact of nonlocal effects due to the metal as a function of the graphene-metal separation, and discuss its implications on the field distribution and plasmonic losses. Finally, we consider a prospective system in which the graphene sheet is replaced by a graphene diffraction grating made of a periodic array of graphene ribbons. In this case, we compute the system's response in the far-field and determine the influence of the metal's nonlocality on the measured spectra. We take realistic parameters for the dielectric functions of the materials that constitute each system, which allows us to compare our results directly to the experiments.


\section{\label{sec:math}Mathematical Details}


Throughout this work, we consider nonmagnetic ($\mu=1$) media. The system under study 
consists in a multi-layered heterostructure, with either dielectric or metallic layers stacked along the $y$--direction, as portrayed in Fig.~\ref{fig:scheme:4layer}. The structure is further assumed to be infinite in the $xz$--plane. We look for TM-modes with a harmonic time dependence in the form of $\ee^{-\ii \omega t}$, and thus the corresponding magnetic field may be written as
\begin{equation}
 \mathbf{H}(\mathbf{r},t) = H^z(x,y) \ee^{-\ii \omega t} \mathbf{\hat{z}}.
 \label{eq:B:general}
\end{equation} 
The form of $H^z(x,y)$ 
follows from Maxwell's equations, and, in general, is different of the regions defined in Fig.~\ref{fig:scheme:4layer}. 
As we consider nonlocal effects only in graphene and in the metal, it is useful to distinguish the description of the fields in the dielectric(s) and metal regions.


\subsection{Dielectric Regions}

In the dielectric media (source-free)
the magnetic field obeys Helmholtz's equation, $\nabla^2 \vect{H} + \vep k_0^2 \vect{H} = 0$, where $k_0=\omega/c$. Assuming a magnetic field in the form of Eq.~\eqref{eq:B:general}, the wave equation admits the following (transverse) solution:
 \begin{equation}
  H^z_\ind{T}(x,y) = \left( C^+ \ee^{\ii k_\ind{T} y} + C^- \ee^{-\ii k_\ind{T} y} \right) \ee^{\ii q x},
  \label{eq:H:dielectric}
 \end{equation} 
where the in-plane wave vector, $q$ (assumed to be along the $x$--direction without loss of generality), and the perpendicular wave vector, $k_\ind{T}$, are related by
 \begin{equation}
  k_\ind{T}=\sqrt{\vep k_0^2 - q^2}.
  \label{eq:kT}
 \end{equation} 
The `T' subscript was introduced to make explicit the transverse nature of these solutions. For uniaxial media (characterized by different permittivities in the $xz$--plane, $\vep^x$, and in the $y$--direction, $\vep^y$), such as hBN, $q$ and $k_\ind{T}$ are connect through an alternative condition, i.e., $k_\ind{T} = \sqrt{\vep^x \omega^2/c^2 - q^2 \vep^x/\vep^y}$.
Maxwell's equation 
$\vect{E}_\ind{T} = (-\ii \omega \vep_0 \vep)^{-1} \vect{\nabla} \times \vect{H}_\ind{T}$ 
enables us to write the corresponding electric field components as
\begin{equation}
 E^x_\ind{T}(x,y) = \frac{- k_\ind{T}}{\omega \vep_0 \vep^x} \left( C^+ \ee^{\ii k_\ind{T} y} - C^- \ee^{-\ii k_\ind{T} y} \right) \ee^{\ii q x},
 \label{eq:Ex:dielectric}
\end{equation} 
\begin{equation}
 E^y_\ind{T}(x,y) = \frac{ q }{\omega \vep_0 \vep^y} \left( C^+ \ee^{\ii k_\ind{T} y} + C^- \ee^{-\ii k_\ind{T} y} \right) \ee^{\ii q x}.
 \label{eq:Ey:dielectric}
\end{equation} 
Naturally, if the dielectric is isotropic, then $\vep\equiv\vep^x=\vep^y$.
 
\subsection{Metal Region}

Within the metal, we assume a Drude dielectric function $\vep_\ind{m}$ of the form
\begin{equation}
 \vep_\ind{m}(\omega) = \vep_{\infty} - \frac{\omega_\ind{p}^2}{\omega^2 + \ii \gamma \omega},
\end{equation} 
 where $\omega_\ind{p}$ and $\gamma$ are the plasma and damping frequencies, respectively, and $\vep_\ind{\infty}$ a background permittivity to account for interband polarization effects. 
 We take nonlocal effects in the metal within the framework of the hydrodynamic model---see  Refs.~\citenum{Raza:2015,moreau2013impact} for details. When taking nonlocality into account, 
Amp\`{e}re's law becomes~\cite{Raza:2015,moreau2013impact}
 \begin{equation}
  \vect{\nabla} \times \vect{H} = - \ii \omega \vep_0 \vep_\ind{m} \left[ \vect{E} - \xi \vect{\nabla}\left( \vect{\nabla \cdot \vect{E}} \right) \right],
  \label{eq:Ampere}
 \end{equation}
 with the parameter $\xi$ defined as
 \begin{equation}
  \xi = \beta\sqrt{\omega_\ind{p}^2/\vep_{\infty} - \omega^2 - \ii \gamma \omega},
 \end{equation}
where $\beta$ is a nonlocal parameter proportional to the Fermi velocity $v_\ind{F}$ of 
electrons in the metal. In this work, we will take the most usual definition\cite{Raza:2015} $\beta = \sqrt{3/5} v_\ind{F}$.
Naturally, the local limit is recovered upon setting $\beta=0$.
 
As discussed above, Maxwell's equations can be shown to admit two kinds of solutions~\cite{Raza:2015,moreau2013impact}: divergence-free and curl-free fields. The former are the usual transverse waves that exist in the local regime.
Within the LRA, the electromagnetic fields in the metal are then described similarly to the fields in the dielectric case, upon replacing $\vep \rightarrow \vep_\ind{m}(\omega)$---see Eqs.~\eqref{eq:H:dielectric}--\eqref{eq:Ey:dielectric}.
On the other hand, curl-free waves do not have an associated magnetic field, as imposed by Faraday's law. However, unlike the local case, the electric field has a non-trivial solution given by the vanishing of the term in square parenthesis figuring in Eq. \eqref{eq:Ampere}, equivalent to the wave equation $\vect{\nabla}^2 \vect{E} - (1/\xi)\vect{E} = 0$. This equation describes longitudinal solutions of the form $\vect{E}_\ind{L} = (E^x_\ind{L} \vers{x} + E^y_\ind{L} \vers{y})\ee^{-\ii \omega t}$, with
\begin{equation}
 E^x_\ind{L}(x,y) =  \left( D^+ \ee^{\ii k_\ind{L} y} + D^- \ee^{-\ii k_\ind{L} y} \right) \ee^{\ii q x},
 \label{eq:Ex:metal}
\end{equation} 
\begin{equation}
 E^y_\ind{L}(x,y) = \frac{ k_\ind{L} }{q} \left( D^+ \ee^{\ii k_\ind{L} y} - D^- \ee^{-\ii k_\ind{L} y} \right) \ee^{\ii q x},
 \label{eq:Ey:metal}
\end{equation} 
with (longitudinal) wave vector
 \begin{equation}
  k_\ind{L} = \sqrt{ - \xi^{-2} - q^2 } = \beta^{-1}
  \sqrt{ \omega^2 + \ii \gamma \omega - \omega_\ind{p}^2/\vep_{\infty}  - q^2 }.
  \label{eq:kL}
 \end{equation} 
Furthermore, $E^x_\ind{L}$ and $E^y_\ind{L}$ are related by the curl-free condition, $\partial E^y_\ind{L}/\partial x = \partial E^x_\ind{L}/\partial y = \ii q E^x_\ind{L}$.
 
The general solution for the fields inside the metal is therefore $H^z = H^z_\ind{T}$ and $E^{x/y} = E^{x/y}_\ind{T} + E^{x/y}_\ind{L}$.
 
\subsection{Boundary Conditions}
 
The coefficients which characterize the fields in each layer are determined by the boundary conditions (BCs). For bound modes (like surface plasmons) we have $q > \sqrt{\vep}\omega/c$. This renders $k$ imaginary, and the signal must be chosen judiciously to ensure that the fields satisfy Sommerfeld's radiation condition. 
For an interface between two dielectrics, the usual BCs apply, that is, the continuity of the tangential component of the electric field ($E^x$) and the (dis)continuity of the magnetic field ($H^z$) in the (presence) absence of a surface current density at the interface. The presence of a finite surface current density is needed in order to describe a graphene sheet (or any other 2D material) placed at an interface between the two media 
, and enters through Ohm's law 
 $\mathbf{K}_\ind{s}(q,\omega) =\sigma(q,\omega) E_{x}(q,\omega) \mathbf{\hat{x}}$
; see also Eq.~(\ref{eq_nonlocal_real_space_sigma}).
Note that here $\sigma(q,\omega)$ entails both frequency and momentum dependencies 
in order to account for nonlocal effects in graphene. We employ graphene's nonlocal 
conductivity using Mermin's particle-conserving prescription, which is detailed in Appendix~\ref{sec:Mermin}.
 
Finally, for a dielectric/metal interface (without graphene~\footnote{We do not consider in this work the case of a dielectric/metal interface separated by a graphene sheet. The main reason is that, under those conditions, doping the graphene in order to adjust its Fermi level would be impractical, since the metal would drain the free electrons in the graphene sheet.}), although the BCs described above remain valid, these need to be augmented by an additional BC (due to the existence of a longitudinal mode within the metal). Typically, this additional BC 
dictates that the normal component of the polarization vector vanishes at metal's surface, given by~\cite{moreau2013impact}
 \begin{equation}
  \mathbf{P} = (\ii/\omega) \vect{\nabla} \times \vect{H} - \vep_0 \vep_{\infty}  \vect{E}.
  \label{eq:polarization}
 \end{equation} 
Note that this polarization field refers to the one associated with the free electrons in the surface of the metal, responsible for the transport of electric currents. Therefore, from a physical perspective, this condition merely imposes that cannot exist currents flowing from the metal to the neighboring dielectric regions (e.g., the interface is a hard wall).
In possession of this additional BC the amplitudes of the fields may now be determined; 
see Appendices \ref{sec:SPP} and \ref{sec:Optical:Calcs} for a more detailed account.

\section{Nonlocal Effects in the Plasmonic Properties} 
 
We consider a system composed by a thin metallic layer (with thickness $h$) 
with a graphene sheet lying at a distance $s$ from its surface. In the spacer region, i.e.,  between graphene and the metal, we assume to have slab of hBN; however, our formalism is general and thus allows for the consideration of different dielectric media. Below the metal and above the graphene, we assume to have air. For the sake of clarity, the system has been divided into four regions, I--IV, as depicted in Fig.~\ref{fig:scheme:4layer}. The calculation of the nonlocal plasmonic properties of the system follows the guidelines discussed in the previous section, and can be consulted with a greater degree of detail in the Appendix~\ref{sec:SPP}. 
In order to assess the influence of the metal's nonlocal effects in the plasmonic properties of the system, we compare our nonlocal results with the corresponding predictions of the local-response theory.

 \subsection{Nonlocal Effects in the Plasmon Dispersion}
 
For the study of the metal's nonlocal effects in the dispersion relation of graphene plasmons, we considered here two different metals---Gold (Au) and Titanium (Ti)---described by the parameters presented in Table \ref{tab:parameters}. The dielectric function of hBN, on the other hand, is adopted from Refs.~\citenum{woessner2015highly} (out-of-plane direction) and \citenum{brar2014hybrid} (in-plane direction).
 \newcommand{\mc}[3]{\multicolumn{#1}{#2}{#3}}
\renewcommand{\arraystretch}{1.2}
\begin{table}[htbp]
\begin{center}
\caption{Drude model parameters used for Titanium\footnote{Note that the authors of Ref. \citenum{rakic1998optical} described Titanium using both Drude and Lorentz terms, but, for the purpose of this work, we only considered the Drude contribution, what nonetheless provides a very good approximation.} (Ti) and Gold (Au). The superscripts indicate the corresponding references.}
\label{tab:parameters}
\begin{tabular}{>{\centering\arraybackslash}p{2cm}>{\centering\arraybackslash}p{2cm}>{\centering\arraybackslash}p{2cm}}
 & Ti & Au\\\hline\hline
$\omega_\ind{p}\mbox{ [eV]}$ & 2.80 \cite{rakic1998optical} & 8.84 \cite{brandt2008temperature} \\
$\gamma_\ind{m}\mbox{ [meV]}$ & 82.0 \cite{rakic1998optical} & 103.0 \cite{brandt2008temperature} \\
$\vep_{\infty}$ & 2.2 \cite{barchiesi2014fitting} & 9.84 \cite{derkachova2016dielectric} \\
$v_\ind{F}/c$ & $0.00597$ \cite{lehtinen2012evidence} & $0.00464$ \cite{ashcroft1976solid}\\\hline\hline
\end{tabular}
\end{center}
\end{table}
 
\begin{figure}[htbp]
 \centering
 \includegraphics[width=\linewidth]{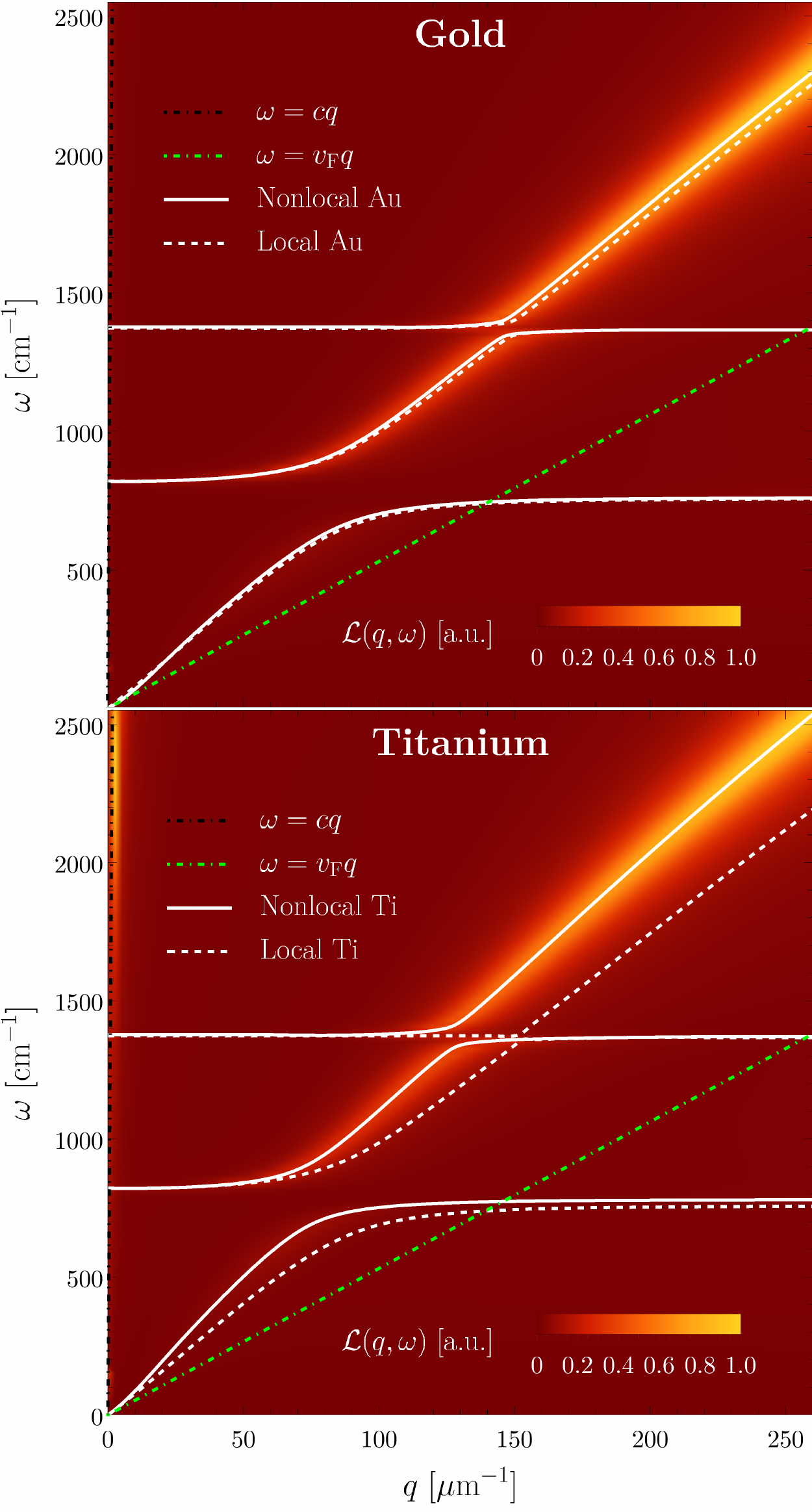}
 \caption{Dispersion relation (DR) of SPPs in an air/metal/hBN/graphene/air configuration, both for a local (dashed white) and nonlocal (solid white) metal, in the case of Gold (top) and Titanium (bottom). The black dot-dashed line corresponds to the light dispersion in vacuum, whereas the green dot-dashed line corresponds to the graphene's electron-hole continuum boundary, $\omega=v_{\mathrm{F}} q$, with $v_\ind{F} \approx c/300$. The color-plot shows the respective loss function, obtained nonlocally (see Appendix \ref{sec:Optical:Calcs}). The remaining parameters are: $s=1\unit{nm}$, $h=10\unit{nm}$, $E_\ind{F}=0.5\unit{eV}$, and $\Gamma=16\unit{meV}$.}
 \label{fig:DR}
\end{figure}

The influence of the nonlocal effects arising from the metal's response is investigated by 
computing the dispersion relation, $\omega(q)$, of graphene plasmons and comparing 
the ensuing spectra obtained assuming a local- and a nonlocal-response.
The outcome is presented in Fig.~\ref{fig:DR} (white curves) for the allowed plasmonic modes in the structure pictured in Fig.~\ref{fig:scheme:4layer}. Since there is dissipation in the system (both in the metal, hBN and graphene), either $q$ or $\omega$ needs to be regarded as a complex quantity in order to fulfill the boundary conditions. In what follows, we have chosen $q$ to be a real number and hence $\omega$ is complex. For the time being, we focus on the real part of the frequency, $\mathrm{Re}\{\omega\}$, and denoting it by $\omega$ for simplicity. The corresponding imaginary part, $\mathrm{Im}\{\omega\}$, associated with the plasmonic losses, will be discussed at a later stage.
 
It is apparent from Fig.~\ref{fig:DR} that the dispersion curves are not continuous, but rather present two asymptotes around $750$ and $1350 \unit{cm^{-1}}$; this feature is shared in both 
local and nonlocal frameworks. These frequencies correspond to the beginning of the hBN's Restrahlen bands, where this material is hyperbolic~\cite{goncalves2016introduction}, and 
are associated with the excitation of surface phonon-polaritons (which in this case hybridize with graphene plasmons)~\cite{goncalves2016introduction}.
 
Moreover, Fig.~\ref{fig:DR} clearly demonstrates that nonlocal effects (associated with the metal) impacts the plasmon dispersion very differently depending on the metal: for Au, the nonlocal and local curves lay very close, whereas in the case of Ti, there is a significant blueshift of the dispersion of graphene plasmons due to the nonlocal effects of the Ti (we stress that graphene is treated as being nonlocal in both cases). Quantitatively, the respective blueshifts are around 2\% and 20\%. We have determined that this difference in the magnitude of the nonlocal effects originates from the cumulative influence of two main factors: the Ti's 
larger nonlocal parameter with respect to Au's ($\beta_\ind{Ti}/\beta_\ind{Au} \simeq 1.29$), and Ti's smaller plasma and relaxation frequencies (specially the former, $\omega_\ind{p,Ti}/\omega_\ind{p,Au} \simeq 0.32$). Both quantities---which are intrinsic to each metal---are crucial for the enhancement of the nonlocal effects in Ti. For this reason, we shall focus on Titanium henceforth in order to illustrate a case in which nonlocal effects are pivotal.

\begin{figure}[htbp]
 \centering
 \includegraphics[width=\linewidth]{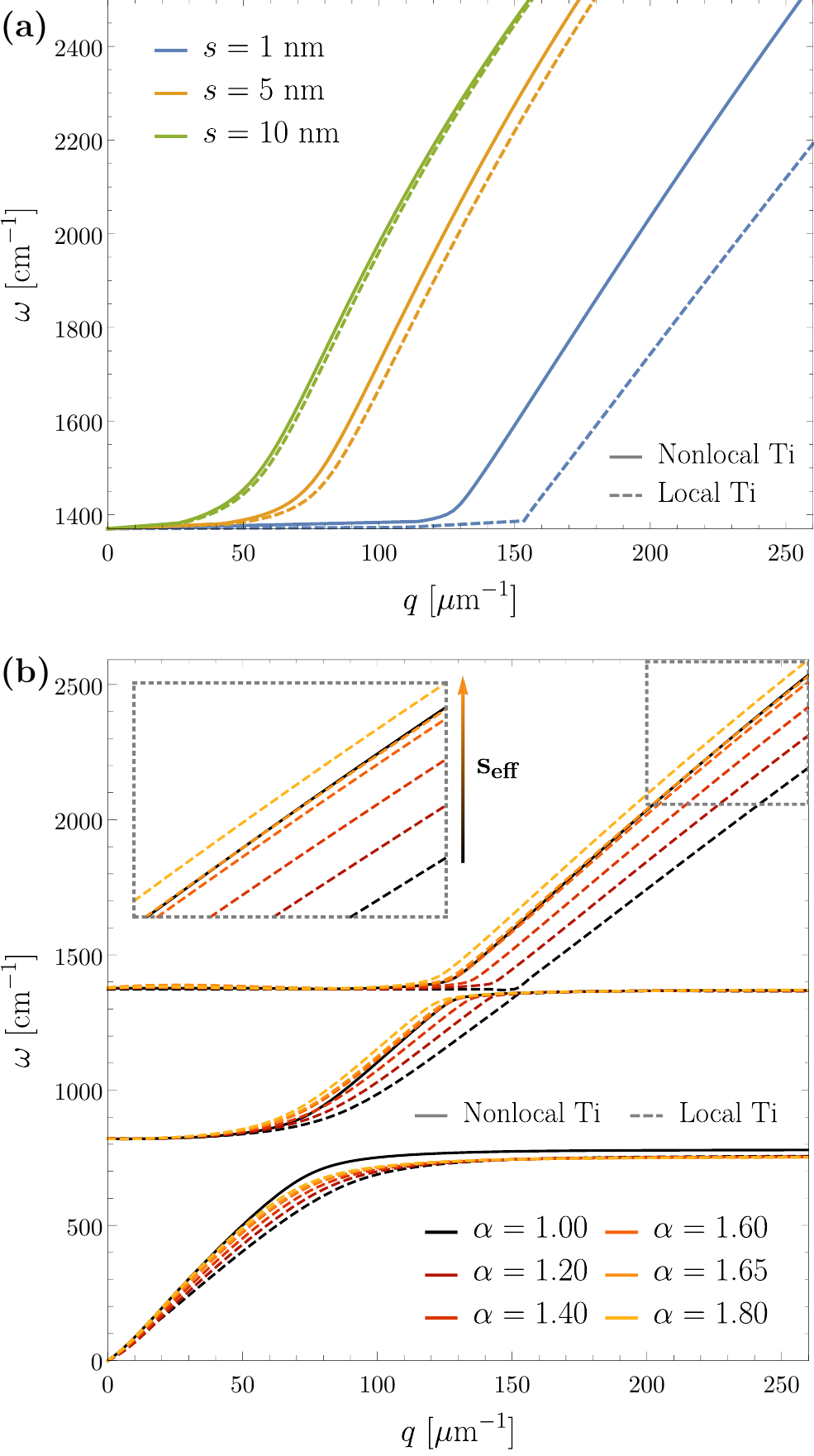}
 \caption{(a) Uppermost branch of the dispersion relation ($\omega>1350$ cm$^{-1}$)  of the plasmonic modes in an air/Ti/hBN/graphene/air configuration, for different values of the hBN's slab thickness, $s$, calculated for a local (dashed) and nonlocal (solid) metal. (b) Dispersion relation in the same configuration, calculated for local and nonlocal metals for different values of the effective parameter $\alpha=s_\ind{eff}/s$, with $s=1\unit{nm}$. The remaining parameters, for both panels, are: $h=10\unit{nm}$, $E_\ind{F}=0.5\unit{eV}$ and $\Gamma=16\unit{meV}$.}
 \label{fig:sVar}
\end{figure}

 It should be emphasized that, apart from the specific characteristics of the metal, the 
 dielectric environment in its vicinity also plays a significant role on the impact of nonlocal effects. In the particular case considered here, the proximity between the metal and graphene (that is, the thickness $s$ of the spacer) is determinant, as Fig.~\ref{fig:sVar}(a) plainly shows. The figure shows the difference between the local and nonlocal calculations increases 
 as the thickness of the spacer region is decreased. In particular, 
while for $s=10\unit{nm}$ the difference is negligible, nonlocal effects become clearly 
perceptible for $s=5\unit{nm}$, and for $s=1\unit{nm}$ nonlocal effects become of paramount importance for an accurate description of the system's plasmonic response. 
This behavior is due to a greater spatial confinement of the fields inside narrower spacers, 
which is enhanced owing to the screening of graphene plasmons by the metal, and in turn gives rise to an acoustic-like graphene plasmons with very high momenta~\cite{lundeberg2017tuning,goncalves2016introduction} and thus more susceptible to nonlocality. 
 
Figure~\ref{fig:sVar}(a) also shows that an increment of the spacer thickness induces a blueshift in the dispersion relation of the plasmonic modes.
For that reason, a na{\"\i}ve approach to account for nonlocal effects while carrying out a local calculation is to consider an \emph{effective} spacer thickness, $s_\ind{eff}$, larger than the actual value, $s$, in a similar fashion to what has been proposed in earlier works~\cite{luo2013surface,luo2014van,derkachova2016dielectric}. Although this method can indeed be regarded as somewhat na{\"\i}ve version of quantum-corrected boundary conditions\cite{Yan:2015,Christensen:2017}, 
it can mimic the proper nonlocal calculation as illustrated in Fig.~\ref{fig:sVar}(b). 
In particular, the nonlocal dispersion is can be reproduced using a local formalism with an effective parameter $\alpha =s_\ind{eff}/s$ between $1.75$ and $1.80$. However, 
it should be stressed that the value of the effective parameter highly depends both on the considered materials (particularly the metal's properties) and on the configuration itself. 
For this reason, this procedure is generally hard to implement because the particular value of $\alpha =s_\ind{eff}/s$ that reproduces the nonlocal effects is difficult to predict theoretically, which renders this approach unsuitable. Instead, it can be merely regarded as a fitting parameter when describing an experiment through local calculations.

\begin{figure}[htbp]
 \centering
 \includegraphics[width=\linewidth]{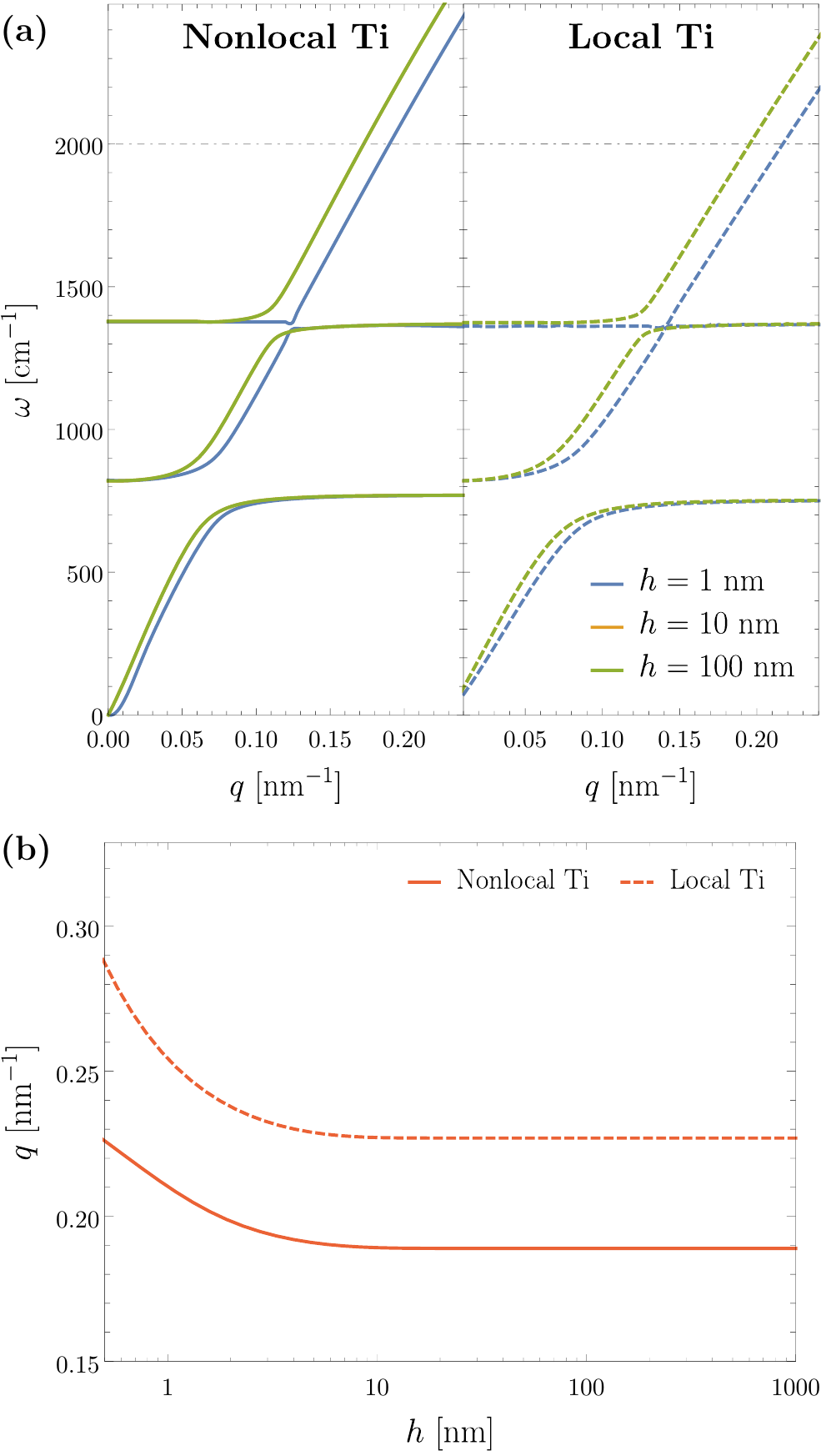}
 \caption{Effect of the variation of the dispersion relation of the plasmonic modes in a air/Ti/hBN/graphene/air structure with the metal thickness, $h$. (a) Nonlocal (left, solid) and local (right, dashed) dispersions for three distinct values of the metal thickness (the orange curve is behind the green one). (b) Plasmon wave vector, for a fixed frequency $\omega=2000\unit{cm^{-1}}$ (dot-dashed on the top panel) as a function of the metal thickness. In both panels, the remaining parameters are: $s=1\unit{nm}$, $E_\ind{F}=0.5\unit{eV}$ and $\Gamma=16\unit{meV}$.}
 \label{fig:hVar}
\end{figure}

Before concluding the present section, we investigate how nonlocal effects vary with the thickness of the metal film. To that end, in Fig.~\ref{fig:hVar}(a) we have plotted the plasmon dispersion relation for an air/Ti/hBN/graphene/air structure using three different values for the Titanium thickness: 1, 10 and 100 nm. The plasmonic spectrum was obtained both nonlocally (left panel) and locally (right panel). In both cases, the graphene is treated as a nonlocal medium. The figure demonstrates that the 10 and 100 nm cases produce the same dispersion, whereas the 1 nm curve lies toward smaller frequencies (for the same $q$). This results suggest that, above a certain threshold thickness, the plasmonic properties of a system with a finite-thickness metal are equivalent to those of a configuration with a semi-infinite metal. This behavior becomes particularly evident upon inspection of Fig.~\ref{fig:hVar}(b), where the plasmon wavevector, for a given frequency, is shown as a function of the metal's thickness. 
Clearly, for $h \gtrsim 3\unit{nm}$, the metallic film is well approximated by a semi-infinite metal (this threshold of $\sim 3\unit{nm}$ is consistent with the penetration depth of the fields in the nonlocal regime, as will be discussed in Section \ref{sec:FieldDist}).

\subsubsection{Comparison to the semi-infinite metal case}

We have seen that for metal thicknesses $h \gtrsim 3\unit{nm}$, the thin-film can be well approximated by a semi-infinite metal. Such a scenario is in fact the most relevant under realistic experimental conditions, which typically employ metals (acting as a gate) with thicknesses in excess of $10\unit{nm}$~\cite{lundeberg2017tuning,alonso2017acoustic}. This 
convenient because it not only simplifies the analysis, but it also allows us to write 
a closed-form expression for the plasmon dispersion in the heterostructure---now a dielectric/graphene/dielectric/metal configuration---, reading
\begin{align}
 \left( \frac{\vep^x_4}{\kappa_\ind{T}^{(4)}} + \frac{\vep^x_3}{\kappa_\ind{T}^{(3)}} + \frac{i \sigma}{\omega \vep_0} \right) \left( 1 + \frac{\vep_{\mathrm m} \kappa_\ind{T}^{(3)}}{\vep^x_3 \kappa_\ind{T}^{\mathrm{m}}} + \delta_\mathrm{nl}  \right) = \nonumber\\
 = 
 \left( \frac{\vep^x_4}{\kappa_\ind{T}^{(4)}} - \frac{\vep^x_3}{\kappa_\ind{T}^{(3)}} + \frac{i \sigma}{\omega \vep_0} \right) \left( -1 + \frac{\vep_{\mathrm m} \kappa_\ind{T}^{(3)}}{\vep^x_3 \kappa_\ind{T}^{\mathrm{m}}} - \delta_\mathrm{nl} \right) e^{-2 \kappa_\ind{T}^{(3)} s}
 \label{res:disRel_D-G-D-M_nonlocal}\ ,
\end{align}
where $\kappa_\ind{T}^{(\nu)} = \sqrt{q^2 \vep_{\nu}^x/\vep_{\nu}^y - \vep_{\nu}^x k_0^2}$ for $\nu=\{3,4\}$, $\kappa_\ind{T}^{\mathrm{m}} = \sqrt{q^2 - \vep_\mathrm{m} k_0^2}$, 
and $\delta_\mathrm{nl}$ is a nonlocal correction term given by~\cite{Raza:2015}
\begin{equation}
 \delta_\mathrm{nl} = \frac{q^2}{\kappa_\ind{L}^{\mathrm{m}} \kappa_\ind{T}^{\mathrm{m}}} \frac{\vep_{\mathrm m}-\vep_\infty}{\vep_\infty}\ ,
\end{equation}
with $\kappa_\ind{L}^{\mathrm{m}} = \sqrt{ q^2 - \left( \omega^2 + \ii \gamma \omega - \omega_\ind{p}^2/\vep_\infty \right)/\beta^2 }$.
Indeed, our calculations demonstrate that Eq.~(\ref{res:disRel_D-G-D-M_nonlocal}) is able to reproduce 
extremely well the plasmon dispersion presented, for instance, in Figs.~\ref{fig:DR} and \ref{fig:sVar}. Furthermore, it is instructive to note that by taking the $\vep_{\mathrm m} \rightarrow \infty$ limit in Eq.~(\ref{res:disRel_D-G-D-M_nonlocal}), which corresponds to the case where the metal becomes a perfect conductor, one obtains (neglecting nonlocal effects)
\begin{equation}
 \frac{\vep^x_4}{\kappa_\ind{T}^{(4)}} \coth\left[\kappa_\ind{T}^{(3)} s \right] 
 + \frac{\vep^x_3}{\kappa_\ind{T}^{(3)}} + \frac{i \sigma}{\omega \vep_0} = 0 \ . 
 \label{eq:DLG}
\end{equation}
Equation~(\ref{eq:DLG}) coincides with the dispersion relation of the acoustic plasmon branch in double-layer 
graphene~\cite{goncalves2016introduction} in a symmetric dielectric environment, where the individual graphene layers are separated by a distance $2s$. This result reflects the scenario in which the screening exerted by the perfect conductor mirrors exactly the charges induced in the graphene sheet. Lastly, 
notice that for large separations, i.e., $s \rightarrow \infty$, the plasmon dispersion~(\ref{res:disRel_D-G-D-M_nonlocal}) reduces to that of an isolated graphene sheet between 
two dielectric media $\vep^x_4$ and $\vep^x_3$.

 \subsection{Nonlocal Effects in the Field Distributions}
 \label{sec:FieldDist}
 
\begin{figure}[htbp]
 \centering
 \includegraphics[width=\linewidth]{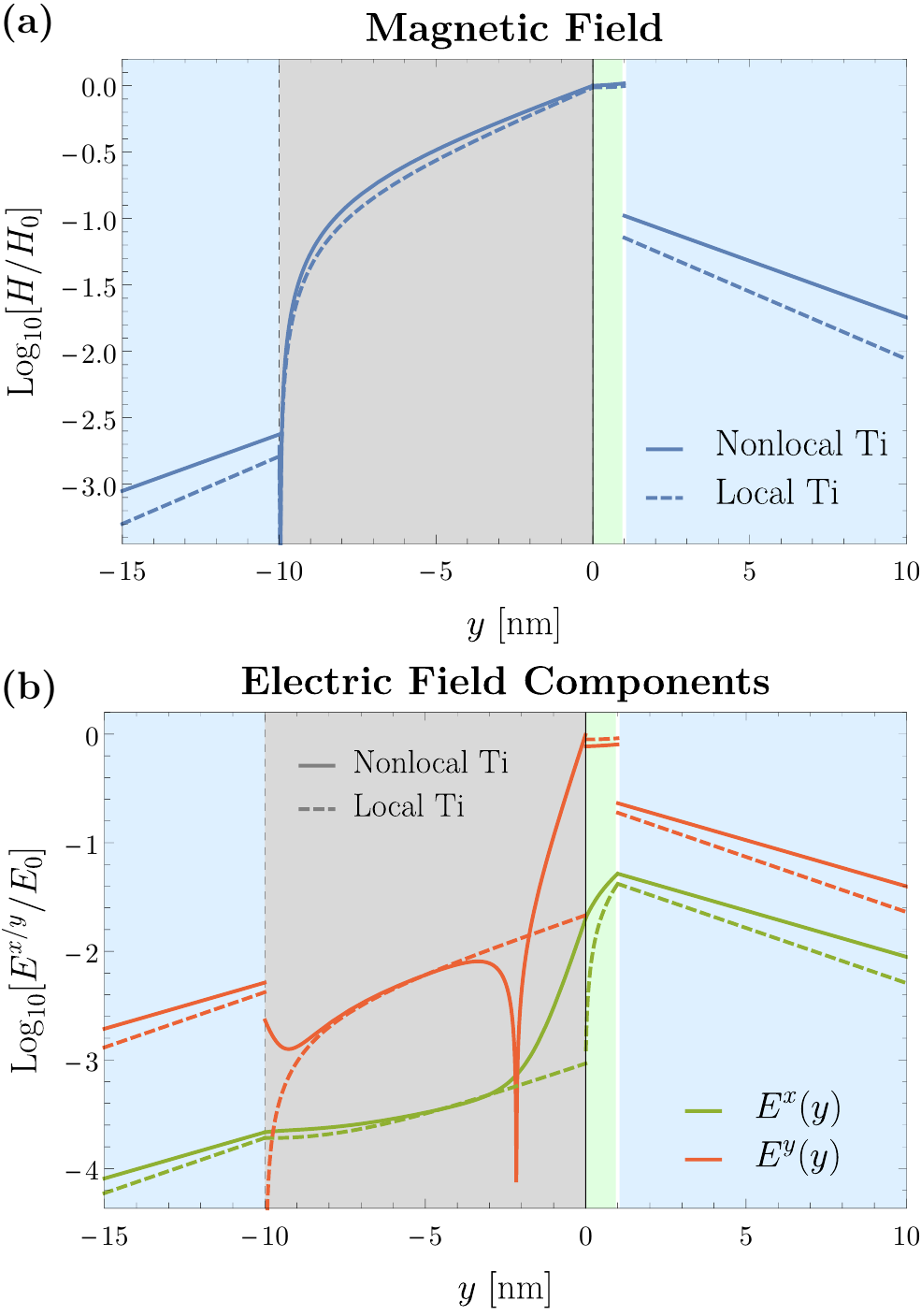}
 \caption{Spacial distribution of (a) the magnetic field and (b) the electric field components, for a air (blue)/Ti (gray)/hBN (green)/graphene (white)/air (blue) structure, calculated using local and nonlocal metal models, for plasmonic modes with the same real part of the frequency $\omega=2000\unit{cm^{-1}}$. $H_0$ and $E_0$ are respectively the nonlocal magnetic and total electric fields calculated at $x=0$. The parameters used are: $h=10\unit{nm}$, $s=1\unit{nm}$, $E_\ind{F}=0.5\unit{eV}$ and $\Gamma=16\unit{meV}$. } 
 \label{fig:HE}
\end{figure}

Having discussed the influence of nonlocal effects in the plasmon dispersion, we now 
study their impact in the spatial distribution of the fields associated with the plasmonic modes. 
The fields amplitudes follow from the BCs, and we determine the $y$--dependence of the 
fields at a given ($q,\omega$)--point which satisfies the plasmon dispersion shown in the previous sectio. In this spirit, Fig.~\ref{fig:HE} depicts the  variation of the absolute value of the magnetic and electric field (in logarithmic units), along the heterostructure, for a (real part of the) frequency $\omega=2000\unit{cm^{-1}}$.
It is apparent from Fig.~\ref{fig:HE}(a) that magnetic field distribution remains essentially unchanged under nonlocal corrections. This is a natural consequence of the hydrodynamic model, 
in which nonlocality only enters in the longitudinal components. Since the magnetic field is 
purely transverse, nonlocal effects do not influence it,
and the only differences between the local and nonlocal cases arise from small differences between the coefficients (and the value of $q$ for the same frequency) that describe the magnetic field on each case.

In contrast to the magnetic field, the inclusion of nonlocal effects in the metal's response 
renders significant changes in the spatial distribution of the electric field, as can 
be seen from Fig.~\ref{fig:HE}(b)--(c). Naturally, the differences between the local and 
nonlocal calculations arise mostly in fields within the metal region, as could be anticipated, 
due to the fact that we have allowed the existence of a longitudinal mode inside the metal, 
in the nonlocal case. 
The figure demonstrates that the nonlocality introduced by this additional longitudinal wave has profound implications in the eletric field's spatial distribution. Specifically, notice that 
the electric field inside the metal in the local case is significantly smaller than the field in the spacer region (as expected for a good metal). However, when nonlocal effects are taken into account, the electric field is increased (when compared with the LRA) by several orders of magnitude in the vicinity of the metal's surface. This feature is consequence of the smearing of the electron density introduced by nonlocality, which translates into a larger penetration 
of the fields inside the metal. 
This penetration depth is practically negligible when taking the local approximation, but becomes important when taking nonlocality into account, as Fig.~\ref{fig:HE} shows. In the latter, the penetration length reaches a couple nanometers, and can become comparable to the metal's thickness for ultra-thin films.

Another feature visible in Fig.~\ref{fig:HE}(b)--(c) is the presence, in the nonlocal case, of 
a sharp dip in the electric field magnitude around $y=-2.5\unit{nm}$, increasing again 
towards the metal/air interface. This is due to a node of its $y$--component, which is caused by a destructive interference between the transversal and longitudinal modes inside the metal. Note that both the transversal and longitudinal modes have $x$- and $y$--components, with the total field being given by $E^{x/y} = E^{x/y}_\ind{T} + E^{x/y}_\ind{L}$; the node occurs when $E^{y}_\ind{T} + E^{y}_\ind{L}=0$. For that reason, across the node there is a change of the sign of the $y$-component of the electric field, dividing the regions where the longitudinal mode amplitude is higher (to the right of the node) or lower (to the left) than the transversal mode amplitude.
 
  

 
 \subsection{Nonlocal Effects on the Plasmonic Losses}
 
\begin{figure}[htbp]
 \centering
 \includegraphics[width=\linewidth]{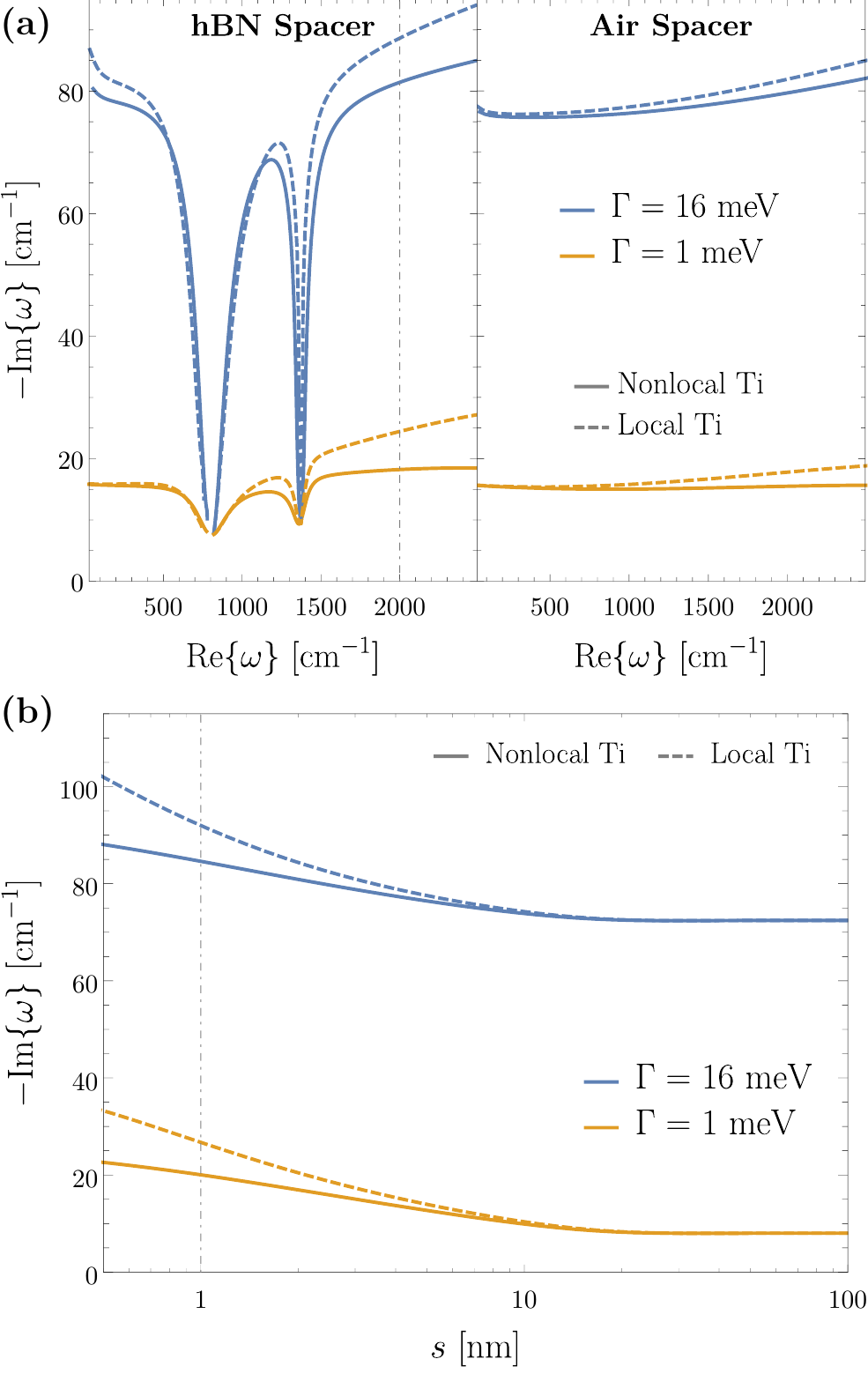}
 \caption{(a) Imaginary part of the frequency of the SPPs in an air/Ti/spacer/graphene/air configuration, with the spacer being hBN (left panel) or air (right panel), calculated using local (dashed) and nonlocal (continuous) models, for two different values of the graphene's electric relaxation energy $\Gamma$, and for $s=1\unit{nm}$ (dot-dashed on the bottom panel). (b) Variation of the imaginary part of the SPPs, in a configuration with a hBN spacer, with the spacer width $s$, for a fixed frequency $\omega=2000\unit{cm^{-1}}$ (dot-dashed on the top panel). The remaining parameters are: $h=10\unit{nm}$ and $E_\ind{F}=0.5\unit{eV}$.}
 \label{fig:Losses}
\end{figure}

We conclude the study of the system portrayed in Fig.~\ref{fig:scheme:4layer} with an analysis on the effect of the nonlocality in the losses associated with the plasmonic modes. 
As mentioned above, in the presence of losses and for a real-valued $q$, the ensuing 
condition for the plasmon dispersion requires a complex-valued $\omega$. So far we have 
limited our discussing to its real part, but here we focus on its imaginary part, 
$\mathrm{Im}\{\omega\}$, since this quantity is intrinsically related with the plasmonic 
losses and plasmon life-time $\tau_\mathrm{p}$, in particular, $\tau_\mathrm{p}^{-1}=-\mathrm{Im}\{\omega\}/2$. %
%
In Fig.~\ref{fig:Losses} we have plotted $-\mathrm{Im}\{\omega\}$ as function of the real part of the polariton frequency, $\mathrm{Re}\{\omega\}$, both within the local and nonlocal response formalism (for the metal; graphene is modeled as nonlocal in all cases).

We have considered two cases, one with a relatively high (for graphene) scattering rate of 
$\Gamma=16\unit{meV}$, and another illustrating high-quality, low-loss graphene, in which $\Gamma=1\unit{meV}$. We note that the change of graphene's electronic scattering rate does not significantly alter the dispersion relation presented in Fig.~\ref{fig:DR} (in fact, there is no visible difference to the eye). However, it significantly impacts the losses affecting 
graphene plasmons sustained in the heterostructure. As Fig.~\ref{fig:Losses} shows, 
the differences between the calculation using a local and a nonlocal metal is modest. This 
suggests that the losses that the plasmons incur originate from the graphene, 
which is natural since the metal only participates indirectly---via screening---and the 
mode's spectral weight is essentially attributed to graphene's.
Furthermore, Fig. \ref{fig:Losses}(b) shows that this difference becomes more evident when the spacer width is reduced, since the losses, when calculated locally, strongly increase for small spacers, whereas nonlocally their increase is very small. This means that, by reducing the spacer size, an increasingly higher confinement can be achieve without a strong increasing in the losses, what is a very important result.

Let us now understand the sudden decrease of the losses seen,  in Fig.
\ref{fig:Losses} [panel (a)] around the hBN phonon frequencies,
when  large damping in graphene (16 meV) is considered. To that end we recall 
that the damping of the phonon modes in hBN are 2.4 meV for the in-plane phonon\cite{brar2014hybrid}
and 1.9 meV for the out-of-plane phonon\cite{woessner2015highly}. At  the frequency of the phonons the polariton spectrum has, essentially, a phononic nature. Therefore, the losses are essentially controlled by those due to phonons. Since these are 
much smaller than 16 meV, we see a sudden decrease in the losses. On the other hand, for a damping in graphene of 1 meV, the phonon damping is comparable to 
the damping in graphene. As consequence the curves of the losses in Fig. \ref{fig:Losses} [panel (a)] do not present the sudden drop at the phonon frequencies. In conclusion, away from the phonon frequencies, the losses are essentially due to  graphene, whereas near the phonon frequencies of hBN the losses are essentially controlled by the behavior of the imaginary part of the dielectric 
function of the hBN spacer. 
 
\section{Probing nonlocality in metals using a graphene nanoribbon grating}

A common approach to experimentally access the plasmonic properties of a system consists on performing reflectance and/or transmittance measurements of the far-field spectra, upon 
illuminating the sample.In this context, plasmon excitations appear as resonances in the 
spectra (as peaks or dips). Therefore, one may investigate the influence of nonlocal effects 
by studying the resulting spectra and compare it to experimental data.
 
In the extended, continuous layered system studied in the previous section, we have seen that the metal's nonlocal response affects the plasmon properties of the modes supported in the system. However, the wave vector mismatch between the graphene plasmons in Fig.~\ref{fig:scheme:4layer} structure and the one of a photon in free-space differs by more than two orders of magnitude, and thus it has been primarily investigated by near-field techniques which are able to overcome this kinematic limitation~\cite{lundeberg2017tuning,alonso2017acoustic}.

\begin{figure}[htbp]
 \centering
 \includegraphics[width=\linewidth]{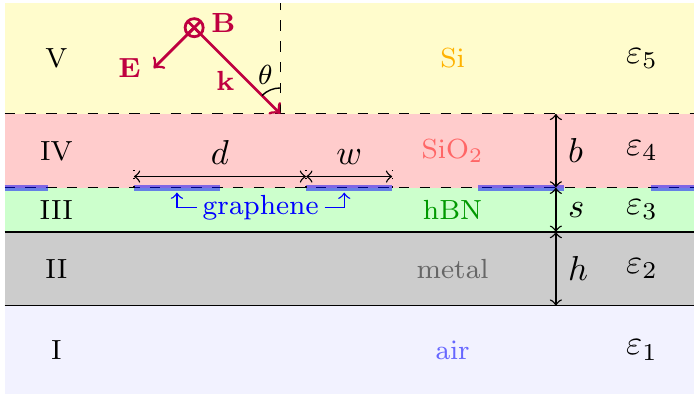}
 \caption{Schematic representation of the layered system considered in this section. It is assumed to be periodic in the $x$-direction and uniform in the $z$-direction. 
 All the different materials, dimensions and regions considered are marked in the figure. Graphene is considered as being nonlocal throughout and the metal is assumed to be either local or nonlocal.}
 \label{fig:scheme:5layer}
\end{figure}

In what follows we consider a different configuration. It is similar to the one considered in the previous section, but we now assume that the graphene monolayer has been patterned into a periodic array of graphene nanoribbons---see Fig.~\ref{fig:scheme:5layer}. These effectively act as a diffraction grating, 
whose Fourier components provide the necessary in-plane momentum to excite plasmons in the 
system~\cite{goncalves2016introduction,gonccalves2016modeling}. Indeed, this mechanism 
not only allows the excitation of graphene plasmons by free-space photons but it also 
serves as a platform susceptible to enhance nonlocal effects, since the higher diffraction 
order can carry substantial momentum and thus promote nonlocal effects. Hence, in order to illustrate a case where nonlocality plays a major role on the optical properties of the system, we will replace the graphene sheet considered in the previous section by a graphene diffraction grating with a period $d$, and ribbon width $w$ (see Fig.~\ref{fig:scheme:5layer}). 
We calculate the optical properties (namely, the reflectance $\mathcal{R}$, transmittance $\mathcal{T}$ and absorbance $\mathcal{A}$) of such structure when it is illuminated by a $p$--polarized plane-wave coming from region V, with monochromatic frequency $\omega$
and incident angle $\theta$, as depicted in Fig.~\ref{fig:scheme:5layer}
Apart from the introduction of the diffraction grating, we will also be considering henceforth a system where the ribbon array is encapsulated between the spacer (hBN, as before) and a thick ($285\unit{nm}$) layer of silicon dioxide (SiO$_2$). The latter's permittivity is taken from Ref.~\citenum{palik1998handbook}. On top of the SiO$_2$, we further consider a semi-infinite layer of silicon (Si) described by a isotropic dielectric constant of $\vep_\ind{Si}=11.66$~\cite{dargys1994handbook}.
Our theoretical calculations for the grating system follow from the considerations outlined in the previous sections and it is based on a Fourier modal expansion of the electromagnetic fields 
described elsewhere~\cite{goncalves2016introduction,gonccalves2016modeling}. In the interest of 
selfcontainedness, we provide the mathematical details in Appendix~\ref{sec:Optical:Calcs}. 
 
 
Figure~\ref{fig:RTA}(a) shows the outcome of our computations for 
the reflectance, $\mathcal{R}$, transmittance, $\mathcal{T}$, and absorbance, $\mathcal{A}$, 
spectrum for a representative 
graphene ribbon diffraction grating placed near a metal, with period $d=25 \unit{nm}$ and ribbon width of $d/2$. Normal incidence ($\theta=0$) is assumed hereafter, but our formalism is 
valid for arbitrary incident angles (though the dependence on the impinging angle is weak since $\sin\theta \ll 2\pi/d$). Furthermore, as before, the impact of the nonlocal effects 
arising from the nearby metal's response are evaluated by comparing the local with the nonlocal case (while maintaining graphene treated at the nonlocal level).
\begin{figure}[htbp]
 \centering
 \includegraphics[width=\linewidth]{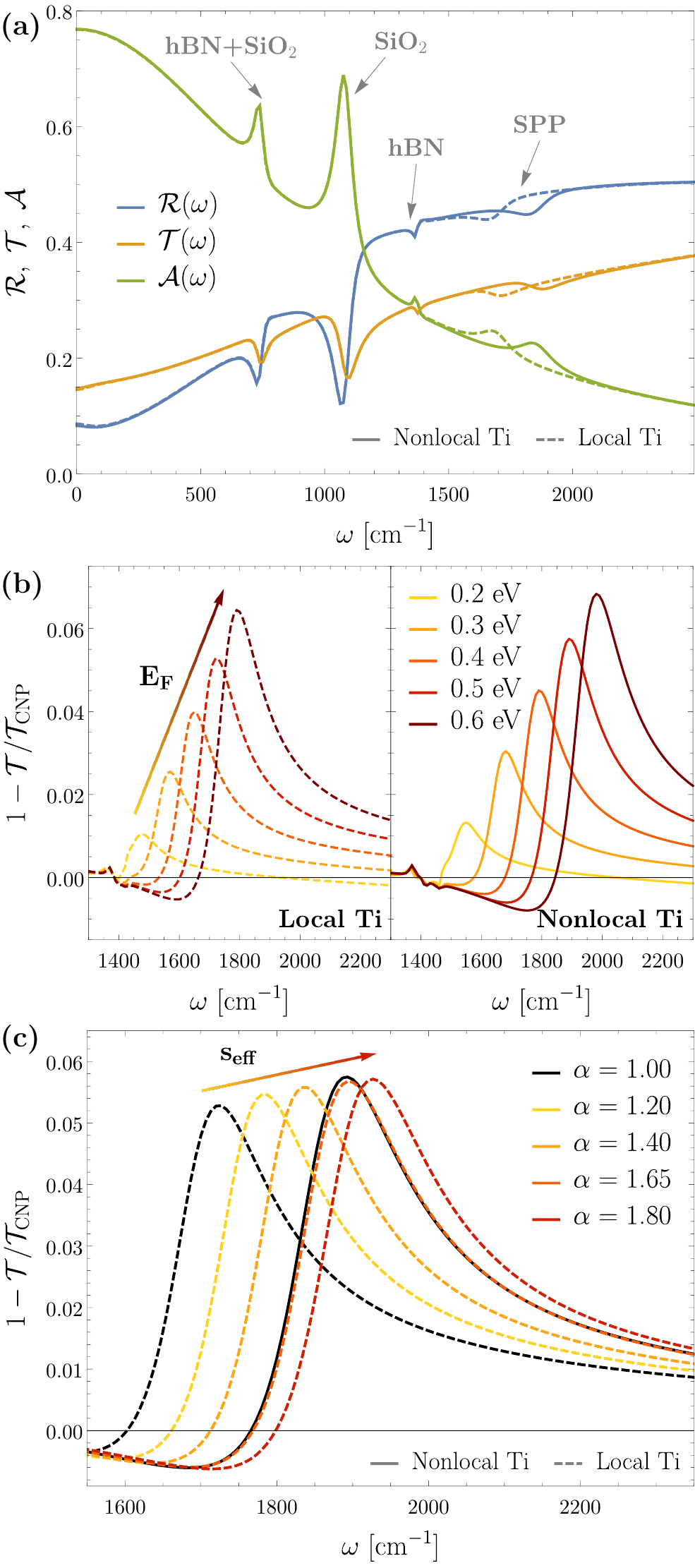}
 \caption{(a) Reflectance ($\mathcal{R}$), transmittance ($\mathcal{T}$), and absorbance 
 ($\mathcal{A}$) spectra of an air/Ti/hBN/graphene/SiO$_2$/Si system. Graphene ($E_\ind{F} = 0.5\unit{eV}$, $\Gamma=8\unit{meV}$) is treated as being nonlocal throughout, whereas the metal is considered both within the local and nonlocal frameworks. The character of each peak (phononic for the first three and plasmonic for the fourth) is indicated by the arrows. (b) Extinction spectra (see details in the main text) near the graphene 
 plasmon resonance frequency for different Fermi levels of graphene. The local and nonlocal cases were plotted separately for clarity. Notice the blueshift of the nonlocal curves relative to their local counterparts. (c) Variation of the extinction curves (with a local description of the metal) of the same system with the effective parameter $\alpha = s_\ind{eff}/s$, compared to the nonlocal counterpart. In all cases, the remaining parameters are: $h=10\unit{nm}$, $s=1\unit{nm}$, $b=285\unit{nm}$, $d=25\unit{nm}$, $w=d/2$ and $\theta=0\dgr$.}
 \label{fig:RTA}
\end{figure} 
The spectral features visible in Fig.~\ref{fig:RTA}(a) are rich, but these can be split 
into two categories, detailed below. The first three peaks/dips---located approximately at $750$, $1100$, and $1350\unit{cm^{-1}}$---do not present any variation due to the nonlocal effects. These peaks correspond to the excitation of optical phonons in either the hBN or the SiO$_2$ slabs (indicated in Fig.~\ref{fig:RTA}(a)). Note that both the hBN and the SiO$_2$ support phononic modes around $750 \unit{cm^{-1}}$.
On the other hand, the position of the fourth resonance shifts considerably to higher frequencies upon inclusion of nonlocal effects in the metal. 
Moreover, in Fig.~\ref{fig:RTA}(b) we have plotted the extinction spectra, defined as $1-\mathcal{T}/\mathcal{T}_\ind{CNP}$ (where $\mathcal{T}$ is the transmittance at some finite Fermi level and $\mathcal{T}_\ind{CNP}$ is the transmittance at the graphene's charge neutrality point), for several values of the Fermi energy. We focus on the spectral window where the graphene plasmon lies. Clearly, the peak disperses towards higher frequencies with increasing 
the Fermi level. This unarguably demonstrates that this resonance corresponds indeed to the excitation of a graphene plasmons in the array. The advantage of exciting these plasmons using a grating is that their frequency is highly dependent on the value chosen for the period and ribbon width of the graphene array, with smaller values of the period yielding concomitant larger wave vectors and thus plasmon resonances at higher frequencies.
 For a Fermi level of $0.5 \unit{eV}$, the difference in the resonance position, between the local and nonlocal metal cases, is roughly $200 \unit{cm^{-1}}$, which in turn corresponds to a relative variation of more than $10 \%$. This plainly shows that, in the case of a graphene sheet placed very close a metallic film, the proper account of nonlocal effects both in the metal and in graphene are key to properly model and interpret experimental results in such a system.
 
 A solution for this problem may be to consider the spacer width $s$ to be an effective fitting parameter, $s_\ind{eff}$, higher than the actual experimental value, as shown in Fig. \ref{fig:RTA}(c).
 
 In Fig. \ref{fig:RTA}(c) is depicted the nonlocal curve for $s=1\unit{nm}$ overlaid by local curves for different values of the effective spacer thickness; for $s_\ind{eff}=1.75 \unit{nm}$, the correspondence between the local and nonlocal curves is nearly perfect, proving the high efficacy of this method.

\section{Conclusions}

Nonlocality in metals is strongly influenced by several material-dependent parameters, namely the plasma frequency, the Fermi velocity, the momentum relaxation rate, and the background dielectric function of the core electrons in the metal. In bulk metals, nonlocal effects can be safely neglected since in that case the wavenumber of the radiation interacting with the metal is, in general, small ($q/k_\ind{F} \ll 1$). However, when the wavenumber is large ($q/k_\ind{F} \sim 1$), as it happens with graphene acoustic plasmons, and the metallic structures are small, compared to the wavelength of the radiation in free space, nonlocality become non-negligible and can change substantially, both qualitatively and quantitatively, the results predicted by a local calculation. Therefore a fully nonlocal calculation is essential when the above conditions are met, if one wants to have a quantitative description of the experimental data. 

Nonlocal effects are particularly strong in Titanium, due to its small plasma frequency, small scattering rate, and large Fermi velocity. Often Titanium is used as an adhesion layer between Gold and other materials, such as hBN; for that reason, nonlocality imparts a significant signature in composite systems when graphene, hBN and Titanium are all brought to close proximity to each other. In this paper, we have shown this to be the case, with the extinction spectrum of such a system showing a clear signature of Titanium nonlocality. The extinction spectrum of the structure presents well defined resonances due to the excitation of surface plasmons in graphene whose spectral position depends strongly on the degree of nonlocality in the metal. Therefore, extinction experiments in a metal/hBN/graphene system provide a viable route to retrieve nonlocal effects in metals, as we have shown here. In addition, nonlocal effects also promote larger propagation lengths of the graphene acoustic plasmons (created by the proximity of the metal) than those predicted by the local theory. We have shown, however, that the local calculation can be rescued if the spacer distance between graphene and the metal is taken as a fitting parameter. Indeed we have shown that an increase of the thickness of the spacer is needed to account for the extinction curves of the structure if one insists in making a local calculation for describing the experimental data. Our work made a detailed analysis of all these aspects and can be used to interpret correctly the optical properties of a graphene and metallic nanostructure in close proximity to each other.

\section*{Acknowledgements}

The authors thank  S\'{e}bastien Nanot and Itai Epstein for valuable discussions and comments. 
E.J.C.D., Yu.V.B. and N.M.R.P. acknowledge support from the European Commission through the project “Graphene-Driven Revolutions in ICT and Beyond” (Ref. No. 785219), and from the Portuguese Foundation for Science and Technology (FCT) in the framework of the Strategic Financing UID/FIS/04650/2013. E.J.C.D. acknowledges FCT for the grant CFUM-BI-14/2016. D.A.I. acknowledges the FPI grant BES-2014-068504. F.H.L.K. acknowledges financial support from the Government of Catalonia trough the SGR grant (2014-SGR-1535), and from the Spanish Ministry of Economy and Competitiveness, through the “Severo Ochoa” Programme for Centres of Excellence in R\&D (SEV-2015-0522), support by Fundacio Cellex Barcelona, CERCA Programme / Generalitat de Catalunya and the Mineco grants Ramón y Cajal (RYC-2012-12281) and Plan Nacional (FIS2013-47161-P and FIS2014-59639-JIN). Furthermore, the research leading to these results has received funding from the European Union Seventh Framework Programme under grant agreement no.696656 Graphene Flagship,  the ERC starting grant (307806, CarbonLight), and project GRASP (FP7-ICT-2013-613024-GRASP). N.~A.~M. is a VILLUM Investigator supported by VILLUM FONDEN (grant No. 16498). Center for Nano Optics is financially supported by the University of Southern Denmark (SDU 2020 funding). Center for Nanostructured Graphene is supported by the Danish National Research Foundation (DNRF103).

\appendix

\section{Calculation of the Plasmonic Properties}
\label{sec:SPP}
 
 In this appendix, we describe the procedure we employed to calculate the dispersion relation and all subsequent analysis of the system presented in Fig. \ref{fig:scheme:4layer}. We will assume, with full generality, that all the dielectric media is described by two dielectric functions along the in-plane ($\vep^x$) and out-of-plane ($\vep^y$) directions.
 
 Following the guidelines discussed in Section \ref{sec:math}, we write the magnetic field in either of the 4 regions that compose the system as
 \begin{align}
  &H^z_\ind{I}(x,y) = \eta \ee^{\kappa_\ind{T}^{(1)} y} \ee^{\ii q x}, 
  \label{eq:H:DR:I} \\
  &H^z_\ind{II}(x,y) = \left( \gamma^+ \ee^{\kappa_\ind{T}^{(2)} y} + \gamma^- \ee^{-\kappa_\ind{T}^{(2)} y} \right) \ee^{\ii q x},
  \label{eq:H:DR:II} \\
  &H^z_\ind{III}(x,y) = \left( \alpha^+ \ee^{\kappa_\ind{T}^{(2)} y} + \alpha^- \ee^{-\kappa_\ind{T}^{(2)} y} \right) \ee^{\ii q x},
  \label{eq:H:DR:III} \\
  &H^z_\ind{IV}(x,y) = \zeta \ee^{-\kappa_\ind{T}^{(4)} y} \ee^{\ii q x},
  \label{eq:H:DR:IV}
 \end{align} 
 where $\eta$, $\gamma^{\pm}$, $\alpha^{\pm}$ and $\zeta$ are undetermined coefficients, $q$ is the in-plane momentum of the modes and $\kappa_\ind{T}^{(\nu)} = \sqrt{q^2 \vep_{\nu}^x/\vep_{\nu}^y - \vep_{\nu}^x \omega^2/c^2}$ is the out-of-plane momentum of the transversal modes.
 
 The electric fields, on the other hand, are given by the expressions
  \begin{align}
  &E^x_\ind{I}(x,y) = \frac{\ii \kappa_\ind{T}^{(1)}}{ \omega \vep_0 \vep_{1}^x } \eta \ee^{\kappa_\ind{T}^{(1)} y} \ee^{\ii q x}, 
  \label{eq:Ex:DR:dielectric:I} \\
  &E^x_\ind{III}(x,y) = \frac{\ii \kappa_\ind{T}^{(3)}}{ \omega \vep_0 \vep_{3}^x } \left( \alpha^+ \ee^{\kappa_\ind{T}^{(3)} y} - \alpha^- \ee^{-\kappa_\ind{T}^{(3)} y} \right) \ee^{\ii q x}, 
  \label{eq:Ex:DR:dielectric:III} \\
  &E^x_\ind{IV}(x,y) = \frac{- \ii \kappa_\ind{T}^{(4)}}{ \omega \vep_0 \vep_{4}^x } \zeta \ee^{-\kappa_\ind{T}^{(4)} y} \ee^{\ii q x}, 
  \label{eq:Ex:DR:dielectric:IV}
 \end{align}  
 for the dielectric regions, and by
  \begin{align}
  E^x_\ind{II}(x,y) = \left[ \frac{\ii \kappa_\ind{T}^{(2)}}{ \omega \vep_0 \vep^x_3 } \left( \gamma^+ \ee^{\kappa_\ind{T}^{(2)} y} - \gamma^- \ee^{-\kappa_\ind{T}^{(2)} y} \right) +  \right. \nonumber \\
  \left. + \left( \delta^+ \ee^{\kappa_\ind{L}^{(2)} y} + \delta^- \ee^{-\kappa_\ind{L}^{(2)} y} \right) \right] \ee^{\ii q x}, 
  \label{eq:Ex:DR:metal}
 \end{align}  
  for the metallic region, with $\delta^{\pm}$ being two additional unknown coefficients that arise from the nonlocality, and $\kappa_\ind{L}^{(2)} = \sqrt{ q^2 - \left( \omega^2 + \ii \gamma \omega - \omega_\ind{p}^2/\vep_\infty \right)/\beta^2 }$ is the vertical momentum of the longitudinal modes. Furthermore, in region II, the normal component of the electric field is given by
  \begin{align}
  E^y_\ind{II}(x,y) = \left[ \frac{q}{ \omega \vep_0 \vep^x_3 } \left( \gamma^+ \ee^{\kappa_\ind{T}^{(2)} y} + \gamma^- \ee^{-\kappa_\ind{T}^{(2)} y} \right) +  \right. \nonumber \\
  \left. + \frac{\kappa_\ind{L}^{(2)}}{\ii q} \left( \delta^+ \ee^{\kappa_\ind{L}^{(2)} y} - \delta^- \ee^{-\kappa_\ind{L}^{(2)} y} \right) \right] \ee^{\ii q x}, 
  \label{eq:Ey:DR:metal}
 \end{align}  
 which fully determines the polarization in the $y$-direction, though Eq. \eqref{eq:polarization}, with the form $P_\ind{II}^y(x,y) = (q/\omega)H^z_\ind{II}(x,y) - \vep_0\vep_\infty E^y_\ind{II}(x,y)$.
 
 There are then 8 undetermined coefficients in the problem: $\eta$, $\gamma^{\pm}$, $\delta^{\pm}$, $\alpha^{\pm}$ and $\zeta$, which are determined through the application of the boundary conditions of the problem:
\begin{align}
 &H^z_\ind{I}(x,-h) = H^z_\ind{II}(x,-h), \label{eq:BC1}\\
 &E^x_\ind{I}(x,-h) = E^x_\ind{II}(x,-h), \label{eq:BC2}\\
 &P^y_\ind{II}(x,-h) = 0, \label{eq:BC3}\\ 
 &H^z_\ind{II}(x,0) = H^z_\ind{III}(x,0), \label{eq:BC4}\\
 &E^x_\ind{II}(x,0) = E^x_\ind{III}(x,0), \label{eq:BC5}\\
 &P^y_\ind{II}(x,0) = 0, \label{eq:BC6}\\ 
 &H^z_\ind{III}(x,s) - H^z_\ind{IV}(x,s) = -\sigma E^x_\ind{IV}(x,s), \label{eq:BC7}\\ 
 &E^x_\ind{III}(x,s) = E^x_\ind{IV}(x,s), \label{eq:BC8} 
\end{align}
 where $\sigma(q,\omega)$ is the conductivity of the graphene sheet, calculated using Mermin's formula in Appendix \ref{sec:Mermin}.
  
 It is straightforward to check that the 8 equations \eqref{eq:BC1}--\eqref{eq:BC8} form an undetermined system of equations, meaning that it cannot be solved for all the unknown coefficients. In order to calculate the dispersion relation of the allowed plasmonic modes, we need therefore to define $\zeta \equiv H_0$ as a free parameter of the problem, and find the remaining coefficients as a function of this parameter. Using arbitrarily the boundary equations \eqref{eq:BC1}--\eqref{eq:BC7}, we find easily the coefficients $\eta$, $\gamma^{\pm}$, $\delta^{\pm}$ and $\alpha^{\pm}$ normalized to $H_0$; we do not show the actual expressions because they are rather ugly and very unintuitive.
 
 
 The remaining boundary condition \eqref{eq:BC8}, on the other hand, can be solved in order to find the relation between $q$ and $\omega$ which ensures the solubility of the system of equations; it corresponds to the dispersion relation of the allowed plasmonic solution. This equation is given by
  \begin{equation}
  \frac{\kappa_\ind{T}^{(3)}}{ \vep_{3}^x } \left( \frac{\alpha^+}{H_0} \ee^{\kappa_\ind{T}^{(3)} s} - \frac{\alpha^-}{H_0} \ee^{-\kappa_\ind{T}^{(3)} s} \right) =   -\frac{\kappa_\ind{T}^{(4)}}{ \vep_{4}^x } \ee^{-\kappa_\ind{T}^{(4)} s},
 \end{equation}  
 where we need to bear in mind that $\alpha^{\pm}/H_0$ are, at this point, totally determined coefficients (function of both $q$ and $\omega$, in general).
 
 The process described above takes explicitly in consideration the nonlocal effects. The counterpart local dispersion relation, on its turn, can be calculated when explicitly considering $\delta^{\pm}=0$ and ignoring the boundary conditions \eqref{eq:BC3} and \eqref{eq:BC6}, proceeding analogously otherwise.

 \section{Calculation of the Optical Properties}
 \label{sec:Optical:Calcs}
 
  In this appendix, we describe the procedure we employed to calculate the optical properties of the system presented in Fig. \ref{fig:scheme:5layer}. We will be considering that the media in regions III and IV may be axial, and we will thus describe it by two dielectric functions along the in-plane ($\vep^x$) and out-of-plane ($\vep^y$) directions. We limit this generalizations to these regions, because if media I and/or V were axial, it would have consequences in the definitions of the reflectance and the transmittance that are out of the scope of this work.
 
 Let us consider that the impinging light carries a momentum $\vect{k}=k_x \vers{x} - k_y \vers{y}$ with $k_x = k_0 \sin(\theta)$, $k_y = k_0 \cos(\theta)$ and $k_0=\sqrt{\vep_5} \omega/c$. Due to the grating, this field will be diffracted, so the scattered fields need to be described by a combination of several different diffraction modes $n$, written in each region as
  \begin{align}
  &H^z_\ind{I}(x,y) = H_0 \sum_{n} \tau_n \ee^{-\ii k_{\ind{T},n}^{(1)} y} \ee^{\ii \rho_n x}, 
  \label{eq:H:RTA:I} \\
  &H^z_\ind{II}(x,y) = H_0 \sum_{n} \left( \gamma^+_{n} \ee^{\ii k_{\ind{T},n}^{(2)} y} + \gamma^-_{n} \ee^{-\ii k_{\ind{T},n}^{(2)} y} \right) \ee^{\ii \rho_n x},
  \label{eq:H:RTA:II} \\
  &H^z_\ind{III}(x,y) = H_0 \sum_{n} \left( \alpha^+_{n} \ee^{\ii k_{\ind{T},n}^{(3)} y} + \alpha^-_{n} \ee^{-\ii k_{\ind{T},n}^{(3)} y} \right) \ee^{\ii \rho_n x},
  \label{eq:H:RTA:III} \\
  &H^z_\ind{IV}(x,y) = H_0 \sum_{n} \left( \phi^+_{n} \ee^{\ii k_{\ind{T},n}^{(4)} y} + \phi^-_{n} \ee^{-\ii k_{\ind{T},n}^{(4)} y} \right) \ee^{\ii \rho_n x},
  \label{eq:H:RTA:IV} \\
  &H^z_\ind{V}(x,y) = H_0 \ee^{\ii k_x x} \ee^{-\ii k_y y} + H_0 \sum_{n} r_n \ee^{\ii k_{\ind{T},n}^{(5)} y} \ee^{\ii \rho_n x},
  \label{eq:H:RTA:V}
 \end{align}  
 where $\rho_n=k_x+2n\pi/d$ is the in-plane momentum of the $n$th mode, as settled by the Bloch Theorem, and $k_{\ind{T},n}^{(\nu)} = \sqrt{\vep_{\nu}^x \omega^2/c^2 - \rho_n^2 \vep_{\nu}^x/\vep_{\nu}^y}$ is the out-of-plane momentum in the region $\nu$ for the transversal waves.
 
 The electric field, in its turn, is written as
 \begin{equation}
  E^x_\ind{I}(x,y) = H_0 \left( \frac{1}{\omega \vep_0 \vep^x_4} \right) \sum_{n} \tau_n k_{\ind{T},n}^{(1)} \ee^{-\ii k_{\ind{T},n}^{(1)} y} \ee^{\ii \rho_n x}, 
  \label{eq:Ex:RTA:I}
 \end{equation} 
 and analougously for regions III, IV and V, whereas, for region II, it is written as,
  \begin{align}
  E^x_\ind{II}(x,y) = \left[ \frac{- k_{\ind{T},n}^{(2)}}{ \omega \vep_0 \vep^x_3 } \left( \gamma^+ \ee^{\ii k_{\ind{T},n}^{(2)} y} - \gamma^- \ee^{-\ii k_{\ind{T},n}^{(2)} y} \right) +  \right. \nonumber \\
  \left. + \left( \delta^+ \ee^{\ii k_{\ind{L},n}^{(2)} y} + \delta^- \ee^{-\ii k_{\ind{L},n}^{(2)} y} \right) \right] \ee^{\ii \rho_n x}, 
  \label{eq:Ex:RTA:metal}
 \end{align}  
 where $k_{\ind{L},n} = \sqrt{ \left( \omega^2 + \ii \gamma \omega - \omega_\ind{p}^2/\vep_\infty \right)/\beta^2 - \rho_n^2 }$ is the out-of-plane momentum for the longitudinal waves. The y-component $E^y_\ind{II}$ is defined analogously. Note that $k_{\ind{L},n}$ and $k_{\ind{T},n}$ were defined appropriately to describe propagating fields.
 
 There is a total of 10 undetermined coefficients in the problem for each mode $n$ ($\tau_n$, $\gamma_n^{\pm}$, $\delta_n^{\pm}$, $\alpha_n^{\pm}$, $\phi_n^{\pm}$ and $r_n$), which must be determined through the boundary conditions of the system. Boundary conditions \eqref{eq:BC1}--\eqref{eq:BC6} and \eqref{eq:BC8} still hold, to which we need to add two more conditions,
 \begin{align}
 &H^z_\ind{IV}(x,s+b) = H^z_\ind{V}(x,s+b), \label{eq:BC9}\\
 &E^x_\ind{IV}(x,s+b) = E^x_\ind{V}(x,s+b). \label{eq:BC10}
\end{align}
 
 To illustrate the next step, let us consider boundary condition \eqref{eq:BC1}, with the form
 \begin{align}
 H_0 \sum_{n} \tau_n \ee^{\ii k_{\ind{T},n}^{(1)} h} \ee^{\ii \rho_n x} = \nonumber \\
  H_0 \sum_{n} \left( \gamma^+_{n} \ee^{- \ii k_{\ind{T},n}^{(2)} h} + \gamma^-_{n} \ee^{\ii k_{\ind{T},n}^{(2)} h} \right) \ee^{\ii \rho_n x}.  
 \end{align} 
 
 Multiplying the previous equation, on both sides, by $\ee^{-\ii \rho_{\ell} x}$ and integrating the resulting equation in one unit cell of the system ($|x| < d/2$), we obtain integrals with the form $\int_{-d/2}^{d/2} dx \ee^{\ii (\rho_n - \rho_{\ell}) x} = d \delta_{n\ell}$, which then yield
  \begin{align}
   \tau_{\ell} \ee^{\ii k_{\ind{T},\ell}^{(1)} h} = \gamma^+_{\ell} \ee^{- \ii k_{\ind{T},\ell}^{(2)} h} + \gamma^-_{\ell} \ee^{\ii k_{\ind{T},\ell}^{(2)} h}.  
 \end{align} 
 
 Notice that the previous equation, for each mode $\ell$, relates only the coefficients of the same index $\ell$ ---or, in other words, the boundary condition is obeyed by each individual mode that composes the total field, and not only by the total field itself, what strongly simplifies the problem. This property is equally obeyed by the remaining \eqref{eq:BC2}--\eqref{eq:BC6}, \eqref{eq:BC8} and \eqref{eq:BC9}--\eqref{eq:BC10} conditions, meaning that, proceeding as illustrated above, we arrive at 9 equations which relate all the coefficients for some particular mode $\ell$. These form a determined system of equations which allows for the calculation of 9 out of the 10 different coefficients, in function of the remaining one. We will henceforth take the undetermined coefficient to be the $r_{\ell}$; in that case, one finds that all the other coefficients are related to $r_{\ell}$ by a linear relation of the form
  \begin{align}
  &\alpha^{\pm}_{\ell} = \theta^{\pm}_{0} \delta_{\ell 0} + r_{\ell} \psi^{\pm}_{\ell},  \qquad \phi^{\pm}_{\ell} = \lambda^{\pm}_{0} \delta_{\ell 0} + r_{\ell} \chi^{\pm}_{\ell},
  \label{eq:alphafromr} 
 \end{align} 
 and analogously for $\tau_{\ell}$, $\gamma^{\pm}_{\ell}$ and $\delta^{\pm}_{\ell}$.

 The remaining unknown coefficient, $r_{\ell}$, must now be determined using the remaining boundary condition, which corresponds to the discontinuity of the magnetic field across the graphene grating, due to the presence of surface currents in the graphene. This condition, analogous to the condition expressed in Eq. \eqref{eq:BC7}, has the explicit form
 \begin{align}
  \sum_{n} \left( \alpha^+_{n} \ee^{\ii k_{\ind{T},n}^{(3)} s} + \alpha^-_{n} \ee^{-\ii k_{\ind{T},n}^{(3)} s} \right) \ee^{\ii \rho_n x} - \nonumber\\ 
  - \sum_{n} \left( \phi^+_{n} \ee^{\ii k_{\ind{T},n}^{(4)} s} + \phi^-_{n} \ee^{-\ii k_{\ind{T},n}^{(4)} s} \right) \ee^{\ii \rho_n x} = \nonumber\\ 
  \sum_{n} \left( \frac{\sigma_n k_{\ind{T},n}^{(4)}}{\omega \vep_0 \vep_4^x} \right)  \left( \phi^+_{n} \ee^{\ii k_{\ind{T},n}^{(4)} s} - \phi^-_{n} \ee^{-\ii k_{\ind{T},n}^{(4)} s} \right) \ee^{\ii \rho_n x},
 \end{align} 
 where $\sigma_n = \sigma(\rho_n,\omega)$ is the conductivity of the graphene. The big difference between this condition and all the others is that graphene's conductivity is only non-zero for $|x|<w/2$; this means that, when multiplying both sides of the equation by $\ee^{-\ii \rho_{\ell} x}$ and integrating it on the unit cell of the system, we obtain
 \begin{align}
  \tilde{\alpha}^+_{\ell} + \tilde{\alpha}^-_{\ell} - \tilde{\phi}^+_{\ell} - \tilde{\phi}^-_{\ell}  = 
  \sum_{n} \left( \frac{\sigma_n k_{\ind{T},n}^{(4)}}{\omega \vep_0 \vep_4^x} \right)  \left( \tilde{\phi}^+_{n} - \tilde{\phi}^-_{n}  \right) S_{\ell n},
 \end{align}  
 with $\tilde{\alpha}^{\pm}_{n} \equiv \alpha^{\pm}_{n} \ee^{\pm \ii k_{\ind{T},n}^{(3)} s}$ (and equivalently for $\tilde{\phi}_n^{\pm}$), and
 \begin{equation}
  S_{\ell n} = \int_{-w/2}^{w/2}dx \ee^{\ii (\rho_n-\rho_{\ell}) x} = \frac{\sin\left[\pi w (n-\ell)/d \right]}{\pi(n-\ell)}.
 \end{equation} 
 
  This equation, unlike all the others, relate coefficients of all the allowed diffraction orders $n$. To solve it, we employ equations \eqref{eq:alphafromr} and, upon some mathematical manipulation, we arrive at an expression of the form $\sum_{n} M_{\ell n} r_n = - F_{\ell}$, where 
 \begin{equation}
  F_{\ell} = (\tilde{\theta}_0^+ + \tilde{\theta}_0^- - \tilde{\lambda}_0^+ - \tilde{\lambda}_0^-) \delta_{\ell 0} - \left[ \frac{\sigma_0 k_{\ind{T},0}^{(4)}}{\omega \vep_0 \vep_4^x} \right] (\tilde{\lambda}_0^+ - \tilde{\lambda}_0^-) S_{\ell 0},
 \end{equation} 
 \begin{equation}
  M_{\ell n} = (\tilde{\psi}_{n}^+ + \tilde{\psi}_{n}^- - \tilde{\chi}_{n}^+ - \tilde{\chi}_n^-) \delta_{\ell n} - \left[ \frac{\sigma_n k_{\ind{T},n}^{(4)}}{\omega \vep_0 \vep_4^x} \right] (\tilde{\chi}_n^+ - \tilde{\chi}_n^-) S_{\ell n}
 \end{equation} 
 
 This equation may be written in the matrix form as $\mathbb{M}\cdot\mathbb{R} = - \mathbb{F}$, with $\mathbb{M}$ being a square matrix with elements $[\mathbb{M}]_{\ell n} = M_{\ell n}$, and $\mathbb{R}$ and $\mathbb{F}$ being columns with elements $[\mathbb{R}]_{\ell} = r_{\ell}$ and $[\mathbb{F}]_{\ell} = F_{\ell}$. The $r_{\ell}$ are hence determined through the solution of that matrix equation, written for a high-enough matrix dimension to ensure the convergence of the solution. The remaining coefficients are then determined by equations \eqref{eq:alphafromr} and the analogous for $\tau_n$, $\gamma^{\pm}_n$ and $\delta^{\pm}_n$.
 
 
 After this process, we can find the reflectance and transmittance of the system by the equations 
 \begin{equation}
\mathcal{R} = \sum_{n \in \mathrm{PM}} \mathrm{Re}\left[ \frac{k_{\ind{T},n}^{(5)}}{\vep_5} \right] \mathrm{Re}\left[ \frac{\vep_5}{k_y} \right] \left| r_n \right|^2,
\end{equation}  
\begin{equation}
\mathcal{T} = \sum_{n \in \mathrm{PM}} \mathrm{Re}\left[ \frac{k_{\ind{T},n}^{(1)}}{\vep^x_4} \right] \mathrm{Re}\left[ \frac{\vep_5}{k_y} \right] \left| \tau_n \right|^2,
\end{equation} 
 where the summations are performed over the propagating modes (PM). The absorbance, on the other hand, is defined as $\mathcal{A} = 1 - \mathcal{R} - \mathcal{T}$.
 
 
 Once again, the optical properties for the local case are calculating when repeating the same procedure whilst setting $\delta_{n}^{\pm} = 0$ and ignoring boundary conditions 	\eqref{eq:BC3} and \eqref{eq:BC6}.
 
 Apart from the optical properties, it is interesting to note that, from the reflectance amplitudes $r_n$, one can calculate the loss function $L(\omega,q) = - \sum_n \mathrm{Im}[r_n]$ (where $k_x$ is substituted by an arbitrarily changeable momentum $q$), which is an alternative way to calculate the dispersion relation of the allowed bound modes of the problem. This is shown in Fig. \ref{fig:DR}, where the loss function was overlaid by the dispersion relation calculated as detailed in Appendix \ref{sec:SPP}, and the correspondence between the two plots is excellent.

 
 

\section{Graphene's Nonlocal Conductivity}
\label{sec:Mermin}

For the conductivity of the graphene sheet, we have used Mermin's formula\cite{mermin1970lindhard,Das:1975aa}, which includes nonlocal effects. Let $x \equiv q/k_\ind{F}$ and $y \equiv \hbar \omega/E_\ind{F}$ be dimensional variables constructed from $q$ and $\omega$, respectively. $E_\ind{F}$ refers to the graphene's Fermi energy, $k_\ind{F} = E_\ind{F}/(\hbar v_\ind{F})$ is the Fermi momentum ($v_\ind{F} \approx c/300$ is the Fermi speed) and $\Gamma$ is the material's relaxation energy. The formula we have used was retrieved from \citeauthor{goncalves2016introduction}\cite{goncalves2016introduction},
  \begin{equation}
  \sigma(q,\omega) = 4 \ii \sigma_0 \frac{\hbar \omega}{q^2} \chi_{\tau}\left(\frac{q}{k_\ind{F}},\frac{\hbar \omega}{E_\ind{F}}\right),
 \end{equation} 
 with $\sigma_0 \equiv e^2 / (4\hbar)$ and
 \begin{equation}
  \chi_{\tau}(x,y) = \frac{\left(1+\ii \frac{\Gamma}{y E_\ind{F}}\right) \chi_\ind{g}\left(x,y+\ii \frac{\Gamma}{E_\ind{F}}\right) }{1 + \ii \frac{\Gamma}{y E_\ind{F}} \chi_\ind{g}\left(x,y+\ii \frac{\Gamma}{E_\ind{F}}\right) / \chi_\ind{g}(x,0) }.
 \end{equation} 
 
 \begin{figure}[htbp]
 \centering
 \includegraphics[width=0.85\linewidth]{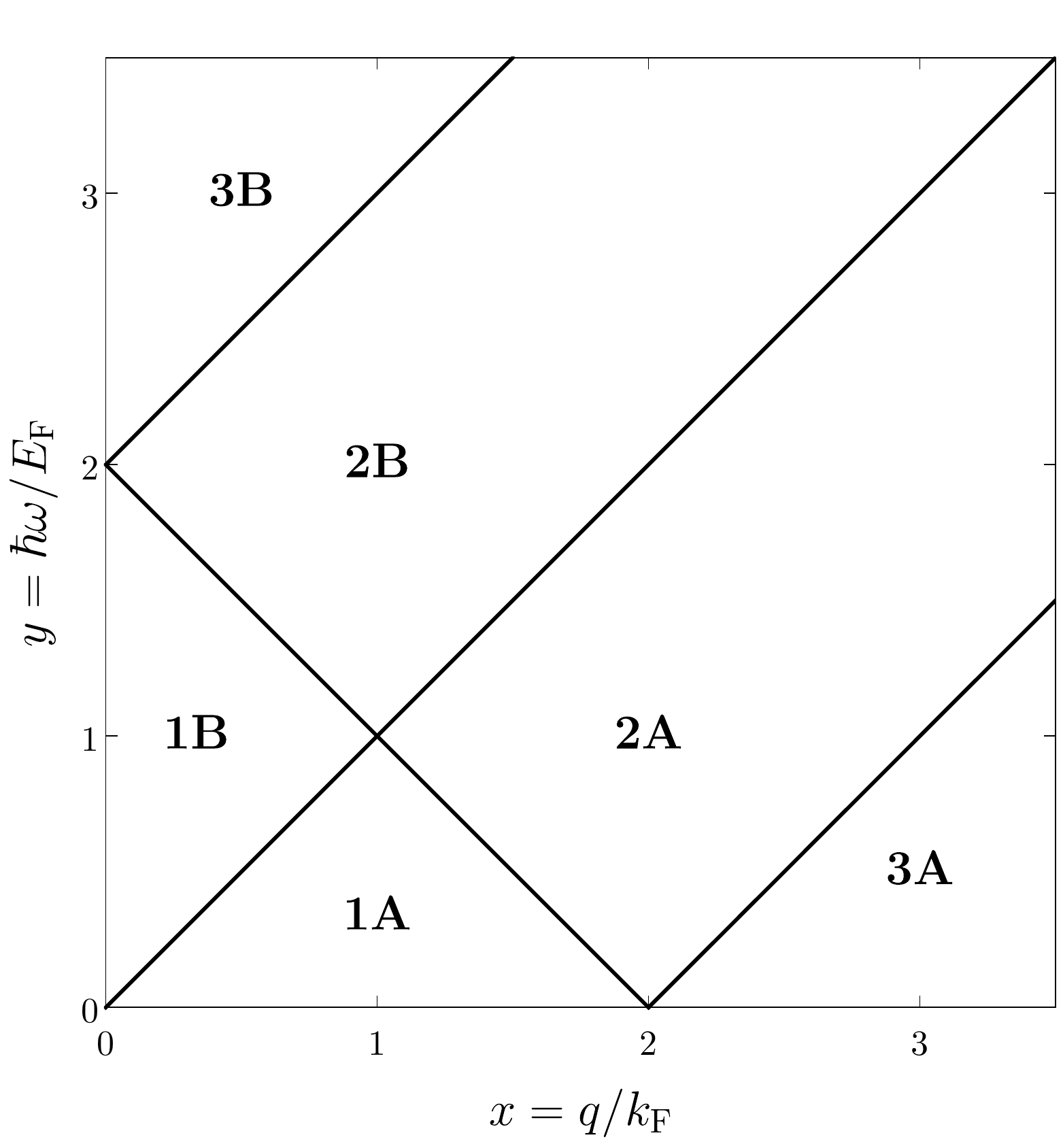}
 \caption{Regions for the calculation of the graphene susceptibility in the $xy$-phase space.}
 \label{fig:mermin}
\end{figure} 
 
 The function $\chi_\ind{g}(x,y)$ is calculated differently according to the region where it is calculated in the $xy$-phase space represented in Fig. \ref{fig:mermin}. It can be written as
 \begin{equation}
  \chi_\ind{g}(x,y) =
  \begin{cases}
   \chi_\ind{B}^\ind{(1)}(x,y), & \ind{Re}[y]>x \wedge \ind{Re}[y]<2-x, \\
   \chi_\ind{B}^\ind{(2)}(x,y), & \ind{Re}[y]>x \wedge \ind{Re}[y]>2-x, \\
   \chi_\ind{B}^\ind{(3)}(x,y), & \ind{Re}[y]>x+2, \\
   \chi_\ind{A}^\ind{(1)}(x,y), & \ind{Re}[y]<x \wedge \ind{Re}[y]<2-x, \\
   \chi_\ind{A}^\ind{(2)}(x,y), & \ind{Re}[y]<x \wedge \ind{Re}[y]>2-x, \\
   \chi_\ind{A}^\ind{(3)}(x,y), & \ind{Re}[y]<x-2,
  \end{cases}
 \end{equation} 
 where $\ind{Re}[y]$ stands for the real part of $y$, and each of the functions in the previous expression are given by
 \begin{align}
  \chi_{\ind{B}}^{\ind{(1)}}(x,y) = -\frac{2}{\pi} \frac{E_{\ind{F}}}{(\hbar v_{\ind{F}})^2} + \frac{1}{4 \pi} \frac{E_{\ind{F}}}{(\hbar v_{\ind{F}})^2} \frac{x^2}{\sqrt{y^2 - x^2}}  \cdot \nonumber \\ \cdot \left[ F\left(\frac{y+2}{x}\right) - F \left( \frac{2-y}{x} \right) \right],
 \end{align} 
 \begin{align}
  \chi_{\ind{B}}^{{\ind{(2)}}}(x,y) = -\frac{2}{\pi} \frac{E_{\ind{F}}}{(\hbar v_{\ind{F}})^2} + \frac{1}{4 \pi} \frac{E_{\ind{F}}}{(\hbar v_{\ind{F}})^2} \frac{x^2}{\sqrt{y^2-x^2}}  \cdot \nonumber \\ \cdot \left[ F\left(\frac{y+2}{x}\right) + \ii G\left(\frac{2-y}{x}\right)\right],
 \end{align} 
 \begin{align}
  \chi_{\ind{B}}^{\ind{(3)}}(x,y) = -\frac{2}{\pi} \frac{E_{\ind{F}}}{(\hbar v_{\ind{F}})^2} + \frac{1}{4 \pi} \frac{E_{\ind{F}}}{(\hbar v_{\ind{F}})^2} \frac{x^2}{\sqrt{y^2-x^2}}  \cdot \nonumber \\ \cdot \left[ -\ii \pi + F\left(\frac{y+2}{x}\right) -F\left(\frac{y-2}{x}\right) \right],
 \end{align}
 \begin{align}
  \chi_{\ind{A}}^{\ind{(1)}}(x,y) = -\frac{2}{\pi} \frac{E_{\ind{F}}}{(\hbar v_{\ind{F}})^2} - \frac{\ii}{4 \pi} \frac{E_{\ind{F}}}{(\hbar v_{\ind{F}})^2} \frac{x^2}{\sqrt{x^2 - y^2}}  \cdot \nonumber \\ \cdot \left[ F\left(\frac{y+2}{x}\right) - F\left(\frac{2-y}{x}\right) \right],
 \end{align} 
 \begin{align}
  \chi_{\ind{A}}^{\ind{(2)}}(x,y) = -\frac{2}{\pi} \frac{E_{\ind{F}}}{(\hbar v_{\ind{F}})^2} + \frac{\ii}{4 \pi} \frac{E_{\ind{F}}}{(\hbar v_{\ind{F}})^2} \frac{x^2}{\sqrt{x^2 - y^2}}  \cdot \nonumber \\ \cdot \left[ \ii \pi - F\left(\frac{y+2}{x}\right) + \ii G\left(\frac{2-y}{x}\right) \right],
 \end{align} 
 \begin{align}
  \chi_{\ind{A}}^{\ind{(3)}}(x,y) = -\frac{2}{\pi} \frac{E_{\ind{F}}}{(\hbar v_{\ind{F}})^2} + \frac{1}{4 \pi} \frac{E_{\ind{F}}}{(\hbar v_{\ind{F}})^2} \frac{x^2}{\sqrt{x^2-y^2}}  \cdot \nonumber \\ \cdot \left[ -\pi + G\left(\frac{y+2}{x}\right) - G\left(\frac{y-2}{x}\right)\right].
 \end{align}
with the functions $F(x) \equiv x \sqrt{x^2-1}-\arccosh(x)$ and $G(x) \equiv x \sqrt{1-x^2}-\arccosh(x)$.

%

\end{document}